%% file: ms.tex
\documentclass[twocolumn,tighten,times,astrosymb]{aastex631}

\usepackage{
	amsmath,
	amssymb,
	amsthm,
	booktabs,
	epsfig,
	graphicx,
	grffile,
	import,
	morefloats,
	microtype,
	natbib,
	stmaryrd,
	subfigure,
	ulem,
	url,
	verbatim,
	wasysym,
	xcolor,
	xspace,
}
\usepackage{savesym}
\savesymbol{tablenum}
\usepackage{siunitx}
\restoresymbol{SIX}{tablenum}
\usepackage[version=4]{mhchem}

\input{commands.tex}

\shorttitle{The ages of \SI{70}{\um}-dark clouds inferred from carbon chain chemistry}
\shortauthors{Worthen et al.}

\begin{document}

\title{
The Young Ages of \SI{70}{\um}-dark Clumps Inferred from Carbon Chain Chemistry
}

\correspondingauthor{Kadin Worthen}
\email{kworthe1@jhu.edu}

\author[0000-0002-5885-5779]{Kadin Worthen}

\affiliation{William H. Miller III Department of Physics and Astronomy, John’s Hopkins University, 3400 N. Charles Street, Baltimore, MD 21218, USA}
\affiliation{School of Earth and Space Exploration, Arizona State University, Tempe, AZ 85281 USA}
\affiliation{National Radio Astronomy Observatory, 1003 Lopezville Rd, Socorro, NM 87801 USA}
\author[0000-0002-8502-6431]{Brian E. Svoboda}
\affiliation{National Radio Astronomy Observatory, 1003 Lopezville Rd, Socorro, NM 87801 USA}
\author[0000-0001-9436-9471]{David S. Meier}

\affiliation{New Mexico Institute of Mining and Technology, 801 Leroy Pl., Socorro, NM 87801 USA}
\affiliation{National Radio Astronomy Observatory, 1003 Lopezville Rd, Socorro, NM 87801 USA}
\author{Juergen Ott}
\affiliation{National Radio Astronomy Observatory, 1003 Lopezville Rd, Socorro, NM 87801 USA}
\author[0000-0001-7594-8128]{Rachel Friesen}
\affiliation{David A. Dunlap Department of Astronomy $\&$ Astrophysics, University of Toronto, 50 St. George St., Toronto, ON, M5S 3H4, Canada}
\author{Jennifer Patience}
\affiliation{School of Earth and Space Exploration, Arizona State University, Tempe, AZ 85281 USA}

\author{Yancy Shirley}
\affiliation{Steward Observatory,  The University of Arizona, 933 N. Cherry Ave., Tucson, AZ 85721}

\begin{abstract}
The physical conditions of the earliest environment of high-mass star formation are currently poorly understood. To that end, we present observations of the carbon chain molecules \ce{HC5N}, CCS, and \ce{HC7N} in the $22-\SI{25}{GHz}$ band towards 12 high-mass \SI{70}{\micron}-dark clumps (SMDC) with the Jansky Very Large Array (VLA).
We detect \ce{HC5N} and CCS towards 11 of these SMDC sources. 
We calculate column densities and abundances relative to \ce{H2} for \ce{HC5N} and CCS. We do not find any clear \ce{HC7N} detections in the 11 sources individually, but by stacking the \ce{HC7N} spectra, we do detect \ce{HC7N} on average in these sources.
We also calculate the ratio of the column densities of \ce{HC5N} to \ce{HC7N} using the stacked spectra of both species.
We compare our measured abundances of \ce{HC5N} and our measured ratio of \ce{HC5N} to \ce{HC7N} to the UMIST dark cloud chemistry models to constrain an age for the gas assuming a fixed volume density and temperature.
The chemical models favor a chemical evolutionary age less than \SI{1}{Myr} at densities of $n(\ce{H2})\approx\SI{2e4}{cm^{-3}}$.
The consistent carbon-chain detections and young model-derived ages support the conclusion that these 11 \SI{70}{\um}-dark clumps lack high-mass protostars because they are young and not because they are inefficient and incapable of high-mass star formation.
\end{abstract}



\section{Introduction}\label{sec:Introduction}
High-mass stars ($>$8$M_{\odot}$) critically influence the evolution of galaxies and the interstellar medium (ISM), yet the initial conditions of high-mass star formation (HMSF) remain poorly understood \citep{2007prpl.conf..165B,2018ARA&A..56...41M}. The formation of high-mass stars is less well understood than their low-mass counterparts \citep{2018ARA&A..56...41M} and differences in the chemical evolution of low- and high-mass star forming regions remains an active area of research \citep{sanhueza2012,feng16,Tatematsu2017}.
Further observations on the initial physical and chemical conditions of protocluster formation are necessary for a better understanding of HMSF. Observations of the incipient evolutionary stages of high-mass star formation are not easily obtained since high-mass stars are less common than low-mass stars, and their formation is obscured by the dense molecular clouds in which they form \citep{2018ARA&A..56...41M}. High-mass stars are thought to form in cold and dense molecular clouds and appear to mainly form in clusters \citep{2003ARA&A..41...57L,2010ApJ...721..222B}. Thus, our understanding of the initial conditions of HMSF and protocluster evolution is dependent on observing and identifying the physical and chemical properties of the cold dense molecular cloud clumps in which high-mass stars form. 

Of particular interest for studying the initial conditions of HMSF are quiescent clumps that are dense ($n(\ce{H2})=10^3-\SI{e5}{cm^{-3}}$), cold ($T<\SI{20}{K}$), and dark at \SI{70}{\um}, as there is a strong correlation between the luminosity of embedded protostars and \SI{70}{\um} flux \citep{Dunham2008,Feng}. As more \SI{70}{\um} dark clouds (SMDCs) are discovered \citep[e.g.,][]{traficante15,svoboda16}, there becomes a larger sample to study the initial physical and chemical conditions of HMSF. A necessary step to using SMDCs to  study the initial conditions of HMSF is first identifying which clumps have the potential to form high-mass stars.

Large samples of high-mass clumps were identified with recent blind surveys of the Galactic plane for dust continuum emission at millimeter and far-infrared (FIR) wavelengths. More than \num{2e3} starless clump candidates (SCCs, {sources where no indicators of high-mass protostars have been found) were identified in the Bolocam Galactic Plane Survey \citep[BGPS;][]{Rosowlowski10,Aguirre11,Ginsburg13,svoboda16}. \cite{svoboda16} identified SCCs by checking for star formation indicators, including \SI{70}{\um} compact sources, H$_2$O and \ce{CH3OH} masers, and ultra compact H\textsc{ii} regions. While high-mass stars are thought to form within SCCs like those identified in these surveys \citep{2010ApJ...721..222B}, it is unclear whether or not these SCCs will go on to form high-mass stars and thus if they are useful for studying the initial conditions of HMSF \citep{RETES-ROMARO2020}. In this study we investigate the ages of a sample of the highest mass SCCs within 5 kpc that were identified with the Bolocam Galactic Plane Survey \citep{svoboda19} to determine if these sources are old and inefficient at forming high-mass stars, or if they are young and have not yet had sufficient time to form high-mass stars.

One method to investigate this question about the ages and star forming potential of these clumps, is to probe their chemistry \citep{Hirota2009, sanhueza12, 2013ChRv..113.8981S}. Knowledge of the chemical composition of molecular clouds is a powerful diagnostic that can be used to determine the physical conditions and evolutionary stages of HMSF, and with the development of high angular resolution interferometers, we can probe the chemistry of individual cores of high-mass star forming regions \citep{2019ApJ...881...57T,Feng,zhu23}. 

More specifically, carbon chain molecules , including the cyanopolyynes (HC$_n$N, $n=$ 3, 5, 7...), are thought to be good tracers of the evolution of star forming regions \citep{Suzuki1992,Hirota2009}. They are formed by gas phase reactions involving carbon atoms and carbon cations (C$^+$) in young molecular clouds before carbon atoms become locked up in CO molecules \citep{2013ChRv..113.8981S,Aguedez13,2018ApJ...854..133T,Tatematsu2017,Taniguchi24}. The formation of carbon chain molecules is thought to happen prior to when molecular clouds reach chemical equilibrium ($\sim$1 Myr) and there still exists free radicals to form the carbon chains \citep{Yamamoto17}. The production of carbon chain molecules can also be increased by the exposure of CO to UV photons in regions of low extinction, which can lead to more carbon atoms in the gas phase, enabling the formation of carbon chain molecules \citep{Spezzano,Bianchi23}. The exposure of CO molecules to UV radiation could lead to the creation of carbon chain molecules at times greater than 1 Myr, however, this is only expected in regions of low visual extinction ($A_V$=2-3) \citep{Snow06}.

Around low-mass protostars, carbon chain molecules have been observed and are thought to form in these regions via a process called Warm Carbon Chain Chemistry (WCCC) \citep{2013ChRv..113.8981S}. Carbon chain molecules can also form through shock chemistry in outflow regions around low-mass protostars (e.g. \citealt{Mendoza18}.) In cold and dense high-mass clumps, however, carbon chain molecules are thought to form at early times before the formation of high-mass protostars \citep{Sakai08,Yamamoto17}, and have been found in young pre-stellar cores (e.g. \citealt{Mengyao19}). WCCC differs from carbon chain chemistry in cold molecular clouds because WCCC is triggered by the sublimation of \ce{CH4} from dust grain mantles, whereas in cold molecular clouds, carbon chain formation is initialized by chemical reactions involving carbon atoms and ions \citep{2013ChRv..113.8981S}.  

After carbon chain molecules are formed during the early evolutionary stages of high-mass molecular cloud clumps, they decrease in abundance at later stages in evolution due to reactions with atomic and molecular ions such as \ce{C+}, \ce{H+}, \ce{H3+}, and \ce{HCO+}, as well as depletion onto dust grains \citep{2013ChRv..113.8981S,Bianchi23}. Carbon chains are thought to be a so-called ``early time'' species in cold clumps as they have been observed to be abundant in starless cores and depleted in further evolved star-forming clumps \citep{Tatematsu2017,2018ApJ...854..133T}, and are thus abundant before WCCC influences molecular abundances at later evolutionary stages. Therefore, carbon chain molecules can be used to probe the evolutionary stage of clumps and consequently their high-mass star forming potential \citep{Rathborne2008}. To that end, we use the NSF's Karl G. Jansky Very Large Array\footnote{The National Radio Astronomy Observatory is a facility of the National Science Foundation operated under cooperative agreement by Associated Universities, Inc.} (VLA) to investigate the carbon chain chemistry of 12 SCCs that were previously identified via the Bolocam Galactic Plane Survey \citep{svoboda16}.

In this paper, we present observations of the carbon chain molecules \ce{HC5N}, \ce{HC7N}, and \ce{CCS} towards 12 SCCs/SMDCs. We provide a description of the observations and targets in Section \ref{sec:Observations}. We describe our methods in obtaining the spectra in Section \ref{sec:Methods}. We report column densities and abundances and compare our results to dark cloud chemistry models in Section \ref{sec:Results}. We discuss the implications of our results in relation to HMSF in Section \ref{sec:Discussion} and conclude in Section \ref{sec:Conclusion}.

\section{Observations}\label{sec:Observations}

\subsection{Molecular Line Observations}
As part of a VLA \ce{NH3} survey of high-mass Starless Clump Candidates (Project code: 17A-146, PI:Svoboda), we simultaneously observed a suite of carbon chain molecules using the WIDAR correlator \footnote{\url{https://science.nrao.edu/facilities/vla/docs/manuals/oss/widar}}.
The observations were made in D configuration with a synthesized beam of $\theta_\mathrm{syn} \approx 3\farcs4$ ($\sim$\SI{0.1}{pc} at \SI{5}{kpc}) and a field of view of approximately 2.6 arcmin (about \SI{3.5}{pc} at \SI{5}{kpc}). The observations were calibrated using the CASA (v5.1) calibration pipeline. The targets were imaged using the CASA task \texttt{tclean} with the ``multiscale'' deconvolver and automated masking via the ``auto-multithresh'' algorithm. Because the emission of the carbon chain lines are rarely bright enough to be deconvolved directly on a per-channel basis, the images are effectively equivalent to their ``dirty image'' counterparts before \textsc{clean}ing in most cases. The absolute flux calibration of the VLA at K-band typically has uncertainties of 5-10$\%$ \footnote{\url{https://science.nrao.edu/facilities/vla/docs/manuals/oss/performance/fdscale}}.

We used the 8-bit samplers and WIDAR to simultaneously observe several sub-bands of carbon chain species in the K-Band (18–26.5 GHz). The observed lines include HC$_5$N $J=9\shortrightarrow8$, CCS $J_N=2_1\shortrightarrow1_0$, HC$_7$N $J=20\shortrightarrow19$, and HC$_7$N $J=21\shortrightarrow20$.
The rest frequencies and channel width of each of the carbon chain transitions we observed are shown in Table \ref{tab:my_label}.
We hereafter refer to these three molecular species (\ce{HC5N}, CCS, \ce{HC7N}) as carbon chains throughout the rest of this paper.
The sub-band spectral resolution for HC$_5$N, CCS and the $J=20\shortrightarrow19$ transition of HC$_7$N are all approximately \SI{0.40}{\kms} and the spectral resolution for the \ce{HC7N} $J=21\shortrightarrow20$ transition is \SI{0.16}{\kms}.

\begin{table}[!h]

    \centering
    \caption{Observational parameters of VLA carbon chain data.}
    \begin{tabular}{c c c c}
    \hline
        Transition & Rest Frequency & Chan.\ Width \\
        & (\si{GHz}) & (\si{\kms}) & Ref. \\
    \hline
         \ce{HC5N} $J=9\shortrightarrow8$& 23.96390100& 0.40 &1 \\
         \ce{CCS} $J_N=2_1\shortrightarrow1_{0}$ & 22.344030 & 0.42&2 \\
         \ce{HC7N} $J=20\shortrightarrow19$ &22.559915 & 0.42 &3 \\
        \ce{HC7N} $J=21\shortrightarrow20$ & 23.687898 & 0.16&3 \\

        \ce{NH3}\ (1,1) &23.6944955& 0.16 &4\\
    \hline
    \end{tabular}
      \begin{minipage}{8 cm}
    \vspace{0.3cm}
    
     \textbf{Note.} Line parameters are taken from CDMS \cite{CDMS}. (1). \cite{Bizzocchi05}, (2). \cite{Lovas92}, (3). \cite{Bizzocchi04b}, (4). \cite{Kukolich67}
    \end{minipage}

    \label{tab:my_label}
\end{table}

\subsection{Target Properties}
The selection of the target sample of 12 clumps is described in \cite{svoboda19} and is briefly summarized here. The clumps were identified through the combination of the images and catalogs from two dust continuum Galactic Plane surveys: (1) the Bolocam Galactic Plane Survey at 1.1 mm and (2) the \cite{IRDC} infrared dark cloud (IRDC) catalog. These sources were identified to be dark at \SI{70}{\um} by visual inspection in \cite{svoboda16}. The sources were also cross-checked with other indicators of star formation activity including compact \SI{70}{\um} sources, mid-IR color selected YSOs, \ce{H2O} masers, Class II \ce{CH3OH} masers, \SI{4.5}{\um} outflows, and ultra-compact H\textsc{ii} regions and were found to possess none of these star formation indicators \citep{svoboda16}. For an in-depth description of the criteria used to determine that these sources do not possess the listed star formation indicators, see \cite{svoboda16}. The lack of star formation indicators in these clumps does not necessarily mean that these clumps are young ($\lesssim$ 1 Myr) and have not had time to form high-mass stars. These sources could also be old and inefficient/incapable at forming high-mass stars. Our observations of ``early time'' carbon chain species can be used as an independent probe, separate from HMSF indicators, to distinguish between the scenarios that these clumps are young and not had enough time to form high mass stars or that they are old and inefficient at forming high mass stars. These 12 specific targets were chosen because they are the highest mass SCCs within a heliocentric distance of \SI{5}{kpc}. All of these 12 sources except G28539, were subsequently discovered to have CO outflows and other indicators of protostellar activity. The protostars within these clumps were determined to be low-mass protostars \citep{svoboda19}. Because these clumps are no longer classified as SCCs, we refer to them, rather, as \si{70}{\micron} dark clumps (SMDCs).

All of the 12 SMDCs in this sample have masses greater than 400 M$_{\odot}$ \citep{svoboda19}. The mass required for a clump to form a single high-mass star ($> 8 M_{\odot}$) is approximately 320 M$_{\odot}$ \citep[see][\S6.1]{svoboda16}, therefore all of the clumps in the sample, have, in principle, sufficient mass for HMSF. The targets have average kinetic temperatures between $10-17$ K, average central volume number densities between $n(\ce{H2}) = \SI{1e4}{cm^{-3}}$ to $n(\ce{H2}) = \SI{1e5}{cm^{-3}}$ , and a median clump mass of 790 $M_{\odot}$ \citep{svoboda19}. The physical properties of these clumps are shown in Table \ref{tab:source_properties}. Cold, dense clumps that contain low-mass young stellar objects are important for studying the initial conditions of HMSF as they are thought to be in the earliest evolutionary stages of HMSF and future sites of high-mass protostellar objects \citep{Molinari2016,Feng}. 

\begin{table}[!htbp]
    
    \centering
    
    \caption{Source Physical Properties}
    
    \begin{tabular}{c c c c c }
    
    \hline
    
        Source & $d_{\odot}$ (pc) &$\Sigma_{\text{pk}}$ (g cm$^{-2}$) & $M_\text{cl}$ ($M_{\odot}$) & Temp (K)  \\
        \hline
        G22695 & 4450 (190) & 0.0580 (0.013) &930 (110)& 15.9 (0.5)  \\
         G23297& 3480 (281) & 0760 (0.019) &420 (85)& 
 12.4 (0.6) \\
         G23481&3780 (220) & 0.1100 (0.024)&760 (120)& 12.7 (0.2)  \\
         G23605&4800 (240) &0.0370 (0.015) &880 (260)& ..  \\
         G24051& 4490 (210) &0.0790 (0.015) &760 (110)& 12.8 (0.5)  \\
         G28539& 4780 (220) &0.1280 (0.011) &3610 (360)&13.6 (0.2)  \\
         G28565&4680 (200) & 0.0830 (0.019)&910 (220)& 13.1 (0.3) \\
         G29558&4370 (240) &0.0690 (0.014) &590 (86)&  14.3 (0.2)\\
         G29601&4270 (280) &0.0900 (0.018) & 660 (130)&17.1 (0.3) \\
         G30120&3680 (260) &0.0750 (0.031) &820 (160)&15.9 (0.2)  \\
         G30660&4410 (240) & 0.0770 (0.019)& 1380 (360)&13.4 (0.4) \\
         G30912&2980 (250) &0.0990 (0.019) &450 (88)&  12.5 (0.2)\\
         \hline
    \end{tabular}
    \begin{minipage}{8 cm}
    \vspace{0.3cm}
    
     \textbf{Note.} Uncertainties are reported in parentheses. Values are taken from Table 2 in \cite{svoboda19}. Temperatures are derived from \ce{NH3} observations (Svoboda et al. in prep). 
    \end{minipage}
    \label{tab:source_properties}
\end{table}


\section{Methods}\label{sec:Methods}
\begin{figure*}[!htbp]
\centering
\includegraphics[scale=.75]{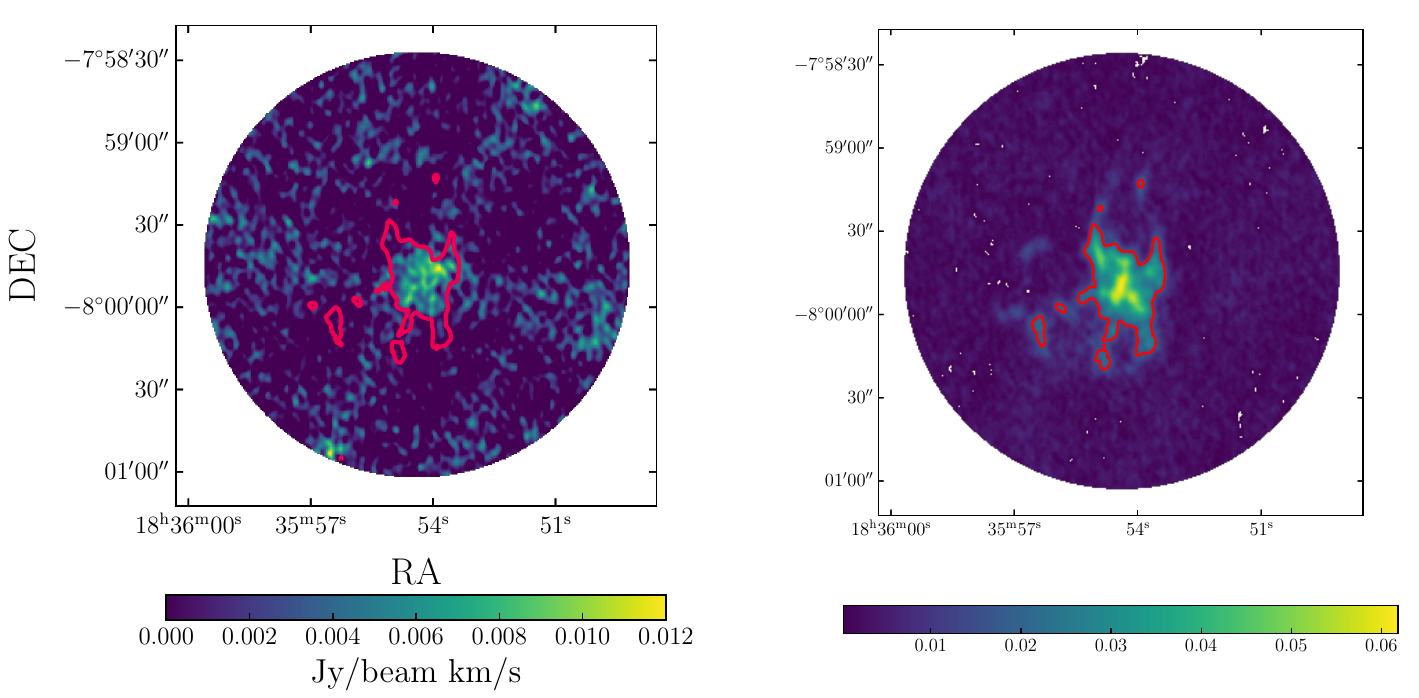}
\caption{Left: moment 0 map of \ce{HC5N} for the source G24051 with \ce{NH3 contours overlaid}. In this source, the \ce{NH3} contour encloses the emission from \ce{HC5N}. Right: moment 0 map of \ce{NH3} (1,1) for the same source as left figure, G24051. The moment calculation was restricted to include only positive values. The red contour line indicates the brightness threshold that was used for the carbon chain extraction aperture in the masking method. }\label{fig:b}
\end{figure*}
We extract the carbon chain spectra from the VLA data cubes using two methods: (1) masking and (2) velocity registration. The spectra for carbon chains are faint, being about 3 times brighter than the single-beam RMS, and spread out over a large area, with the scale of these clumps being around \SI{1}{pc}. The faint emissions from \ce{HC5N} can be seen in the moment 0 map of \ce{HC5N} for the source G24051, which is shown in Figure \ref{fig:b}. The other 10 sources with carbon chain detections have similarly faint and extended carbon chain emission. Because the signal from the carbon chains in these sources has a low signal-to-noise ratio (SNR) per pixel, and is spread out over a wide field, averaging techniques are required to be able to detect the carbon chain spectral lines and precisely measure their integrated fluxes. More specifically, the masked averaging method is necessary to spatially match the signal and improve the SNR of the spectra.

The first averaging technique we use to extract the carbon chain spectra from the VLA data cubes is masking. The carbon chain data cubes were masked based on the moment 0 map of the NH$_3$ (1,1) line such that only pixels above a $3\sigma$ brightness threshold, which corresponds to a value of approximately 0.015 Jy bm$^{-1}$ km s$^{-1}$, on the NH$_3$ (1,1) moment 0 maps for each source, were summed into the carbon chain spectra. We use \ce{NH3} (1,1) as a template for the mask because the \ce{NH3} emission is  brighter than the carbon chains, $\sim\!5$ times brighter in these sources, and from visually comparing moment 0 maps, the carbon chain emission is found within the extent of the \ce{NH3} emission in these clumps (see Figure \ref{fig:b} and the Appendix). Once the pixels below the brightness threshold were masked, the flux from pixels within the $3\sigma$ brightness threshold were summed along the spatial axis. Masking based on the \ce{NH3} moment 0 map improves the peak-SNR by a factor of approximately 1.5 compared to tracing an extraction aperture manually. We also sum the pixels that are outside of the mask (where we do not expect carbon chain emission) to confirm that this does not result in a carbon chain detection. We find that summing the pixels outside of the mask does not result in a 3$\sigma$ detection of \ce{HC5N} or CCS in any of the sources. 

The brightness threshold used for the mask is illustrated as the yellow contour line on the moment 0 map of \ce{NH3} of source G24051 in  Figure \ref{fig:b}. This yellow contour line shows which pixels were masked and which were summed into the spectra for this source. This masking method was used for all 12 of the SMDC sources. Figure \ref{fig:ATLASGAL_nh3} shows the ATLASGAL survey \citep{Schuller,ATLASGAL} \SI{870}{\mu m} images of each of the sources with the \ce{NH3} contours overlaid. In each of the sources, the dust emission region covers the emission region of the \ce{NH3} in the 2D images.

\begin{figure*}
    \centering
    \includegraphics[scale=0.57]{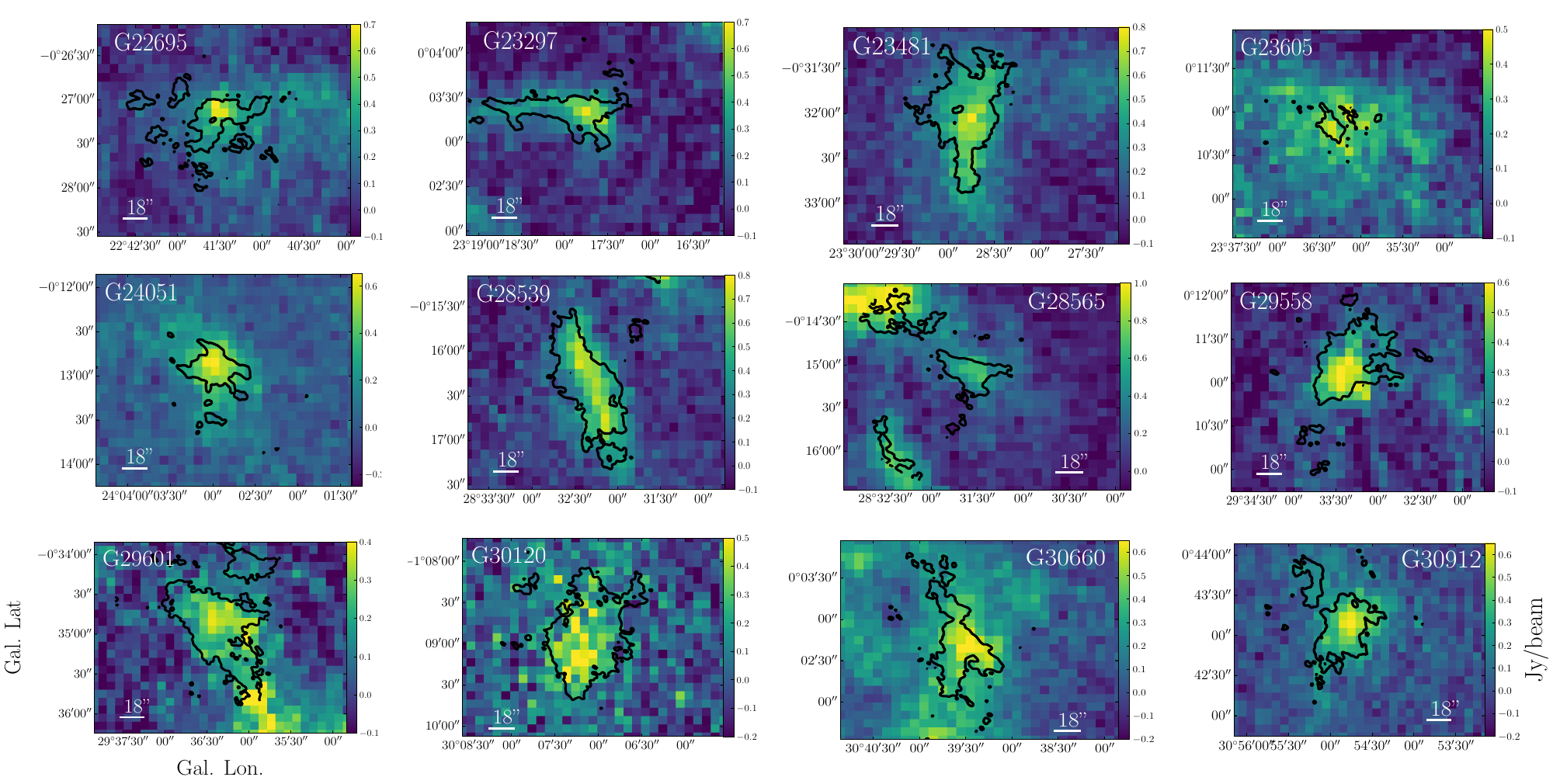}
    \caption{ATLASGAL images at \si{870}{$\mu$m} for all 12 sources with the 3$\sigma$ \ce{NH3} contours overlaid. These contours show the masks used for the extraction of the carbon chain spectra as well as the aperture used to measure the flux from the ATLASGAL images for the calculation of \ce{H2} column density. In each of the sources, the \ce{NH3} and the dust continuum emission appear to be co-spatial. Note that the coordinates for these images are Galactic coordinates, while those in Figure \ref{fig:b} are in RA and DEC and thus the \ce{NH3} contours appear rotated here compared to   Figure \ref{fig:b}.}
    \label{fig:ATLASGAL_nh3}
\end{figure*}

The second method used to obtain the carbon chain spectra was velocity registration. Since these molecular clumps are turbulent and contain spatially varying bulk motions, emission from the carbon chain species occur at different velocities, which dilutes the SNR when spatially averaged. The velocity registration method accounts for the velocity field of the clumps and further improves the peak-SNR of the carbon chain spectra. 

To account for the varying clump kinematics, we register the flux from the carbon chains to the intensity-weighted mean velocity of the \ce{NH3} (1,1) as computed in a moment 1 map for each clump.
This is done by calculating the velocity offset for each pixel based on the value of the moment 1 map and adding that offset to the spectral axis of each pixel. As above in ``method 1,'' values are only computed within the $3\sigma$ mask generated from the moment 0 map. By shifting the spectra of each pixel, the individual spectra are registered so as to be aligned to a consistent system velocity across the clump. Once the spectra of each pixel have been registered, the channels for each pixel are then re-gridded to a common spectral axis using the ``nearest'' interpolation method. Finally, the spectra for each pixel within the spatial mask are summed into an average spectrum. We use the \ce{NH3} moment 0 maps as the extraction apertures for the carbon chains instead of the ATLASGAL images because the \ce{NH3} observations have higher spatial resolution than the ATLASGAL images. The \ce{NH3} data also contains kinematics for each clump, which is used for the velocity registration method. The \ce{NH3} masks, as shown in Figure \ref{fig:ATLASGAL_nh3}, encapsulate the brightest regions of dust emission in each source, so the extracted carbon chain spectra come from the same region as the bulk of the dust emission when using the \ce{NH3} masks as extraction apertures.

For the analysis of the spectra, all were obtained with the masking method alone except the \ce{HC5N} and CCS spectra for G22695, and the CCS spectra for G30120, and G30660. These were obtained using the velocity registration method since these three sources had fainter carbon chain lines than the others and were not clear 3$\sigma$ detections with the masking method alone. Thus, the velocity registration method was required for a carbon chain detection and the measurement of the line flux in these sources. 

We fit the extracted spectra with Gaussians and integrate the best fit Gaussians to determine the total integrated flux of each of the spectral lines. We find that both the masking method and velocity registration method are consistent with each other in that all the carbon chain detections obtained with the masking method were also detected using the velocity registration method. The velocity registration method most likely does not bias the integrated flux measurement of the carbon chain spectral lines, since it is  moving flux to a common velocity based on the \ce{NH3} moment 1 map of the clumps, and is not adding or removing signal or noise from the spectra. We confirm this by measuring the integrated intensity of the carbon chain molecules after applying both methods for sources with clear carbon chain detections using both methods. The total integrated intensity of the spectra obtained using the two methods for these sources were consistent with each other (within the uncertainties). The uncertainties on the amplitude of the line fluxes are measured by calculating the standard deviation of the data points outside of each of the the carbon chain lines. This uncertainty is then propagated with the uncertainty on the line width from the Gaussian fits to determine the uncertainty on the integrated intensity. The total integrated intensity does not depend on the method of spectral extraction that we use, and thus the method of extraction is not expected to bias our column density measurements.

\section{Results}\label{sec:Results}
\subsection{Molecule Detections}
We detect HC$_5$N and CCS in 11 out of 12 of our targets, with G23605 being the only non-detection of either. We consider a species to be detected if we observe an intensity exceeding $3\sigma$ across two or more channels. The brightness of the HC$_5$N and CCS spectra vary over the 11 sources with detections, which can be seen in the spectra of HC$_5$N shown in Figure \ref{fig:d} and the spectra of CCS shown in Figure \ref{fig:e}. Of the sources with carbon chain detections, the brightest HC$_5$N line occurs in the source G30912 and the brightest CCS line is in G28565, while the faintest HC$_5$N lines occurs in G30120 and the faintest CCS line occurs in G30660. 
The results from the Gaussian fits and total integrated intensity calculations for each of the lines can be seen in Table \ref{table:a}.

While we detected \ce{HC5N} and CCS in 11 of our sources, we did not find a clear detection of HC$_7$N in any one of our sources individually. We further checked for an HC$_7$N detection by stacking the HC$_7$N spectra from all of our 11 sources with carbon chain detections after performing the velocity registration method to align all sources at a common velocity of 0 km/s. We average both HC$_7$N transitions we observed, $J=20\shortrightarrow19$ and $J=21\shortrightarrow20$, from all of our sources with carbon chain detections (i.e., excluding G23605). The stacked spectrum for \ce{HC7N} is shown in Figure \ref{fig:f}. We find that we do detect \ce{HC7N} in the stacked spectrum and that the stacking technique does reduce the noise level in the spectrum by approximately a factor of 5. 

To check if the detection of \ce{HC7N} in the stacked spectrum is dominated by a single source we remove one source from the stacked spectrum at a time and measure the signal-to-noise ratio of the \ce{HC7N} detection. The signal-to-noise ratio of the detection does not drop below 5 when removing any of the individual sources from the stacked spectrum, so we conclude that the stacked spectrum is not dominated by a single source.

\begin{table*}[!ht]
 
\centering
 \caption{Line Parameters}
\begin{tabular}{c c c c c c }
   
    \hline
   
    Source & Molecule& T$_B$ & V$_{lsr}$ &FWHM & $\int T_{B} dv $ \\
       &  & (K) & (\si{\kms})& (\si{\kms})& ($\mathrm{K\,km\,s^{-1}}$)\\
    \hline
        G22695 & HC$_5$N & 0.12 (0.03)&77.6 (0.1)& 1.0 (0.3) & 0.13 (0.03)  \\
        G23297 & HC$_5$N & 0.18 (0.02)&55.2 (0.1)& 1.6 (0.2) & 0.31 (0.03) \\
         G23481 & HC$_5$N& 0.07 (0.01) & 63.9 (0.2)&1.8 (0.4) &0.14 (0.03) \\
         G24051 & HC$_5$N& 0.24 (0.02)& 80.9 (0.1)& 1.5 (0.1) & 0.39 (0.03) \\
         G28539& HC$_5$N& 0.08 (0.01)& 88.5 (0.1) & 1.6(0.2)& 0.14 (0.02) \\
         G28565& HC$_5$N& 0.26 (0.03) &87.0 (0.1)&2.0 (0.2) & 0.55 (0.05) \\
         G29558& HC$_5$N & 0.20 (0.02) & 79.9 (0.1)& 1.5 (0.1)&0.32 (0.02) \\
         G29601 & HC$_5$N& 0.06(0.01)& 75.5 (0.4) & 1.5 (0.4)& 0.09 (0.03) \\
         G30120 & HC$_5$N& 0.05 (0.012)& 66.1 (0.2)&1.6 (0.40)& 0.09 (0.01)\\
         G30660 & HC$_5$N& 0.10 (0.01)& 81.0 (0.1)& 1.3 (0.1) & 0.15 (0.01) \\
         G30912 & HC$_5$N& 0.44 (0.05)& 50.5 (0.1)&1.1 (0.1) &0.53 (0.06) \\
    \hline 
         G22695 & CCS & 0.08 (0.03)& 78.0 (0.4)& 2.4 (0.8)& 0.20 (0.06) \\
        G23297 & CCS  &0.32 (0.02) & 54.8 (0.1)& 1.6 (0.1) & 0.55 (0.04)\\
         G23481 & CCS & 0.15 (0.01)& 63.8 (0.1)& 1.9 (0.2) & 0.30 (0.03) \\
         G24051 & CCS & 0.23 (0.02)& 80.9 (0.1)& 1.3 (0.3) & 0.33 (0.03) \\
         G28539& CCS & 0.23 (0.02)& 88.3 (0.1)&1.6 (0.1) & 0.35 (0.03) \\
         G28565& CCS & 0.53 (0.03) &87.0 (0.1)& 1.7 (0.1) & 0.98 (0.05) \\
         G29558& CCS  & 0.22 (0.02) & 79.9 (0.1)& 2.2 (0.2)& 0.53 (0.05) \\
         G29601 & CCS & 0.14 (0.01)& 75.2 (0.1)& 1.8 (0.1) & 0.25 (0.02) \\
         G30120 & CCS & 0.16 (0.02)& 65.2 (0.1)& 2.1 (0.2) & 0.36 (0.04) \\
         G30660 & CCS & 0.16 (0.01)& 80.7 (0.1)&1.4 (0.1) & 0.24 (0.01) \\
         G30912 & CCS & 0.50 (0.06)& 50.5 (0.1)& 0.9 (0.1) & 0.41 (0.06) \\
      \hline
    \end{tabular}
\label{table:a}
 \begin{minipage}{10cm}
    \vspace{0.3cm}
    
     \textbf{Note.} Parameters for the best fit Gaussian to each of the carbon chain lines. The uncertainties are shown in parentheses. 
    \end{minipage}

\end{table*}

\subsection{Column Density Calculations}
Using the total integrated intensity of each line, we calculate the column densities of HC$_5$N and CCS assuming  optically thin lines. For \ce{HC5N}, we use the equation for the column density of the upper level $J_u$ of a linear molecule from \citep{1980A&A....81..245W}:
\begin{equation}
    N_u(J_u)= \num{8.348e16}\frac{(2J_u-1)}{J_u^2}\frac{\int T_Bdv}{\mu^2B} \ \left[\si{cm^{-2}}\right],
\end{equation}
where $J_u$ is the upper level in the transition $J_u \rightarrow J_{u-1}$, $\mu$ is the dipole moment in Debye, $\int T_B dv$ is the integrated intensity of the spectra, and $B$ is the rotational constant in MHz. We use $B=\SI{1331.33}{MHz}$ and $\mu = \SI{4.33}{D}$ for the HC$_5$N $J=9\shortrightarrow8$ transition, which are taken from \cite{1976JMoSp..62..175A}. To calculate the total column density, we use the relation between the column density of a molecule in the upper energy state and the total column density of a molecule, assuming the level populations are in local thermodynamic equilibrium (LTE), which is given as
\begin{equation}
\frac{N_\mathrm{tot}}{N_u}=\frac{Q_\mathrm{rot}}{g_u}e^{\frac{E_u}{kT_\mathrm{ex}}}
\end{equation}
where $g_u$ is the degeneracy of the energy state, which is $2J_u+1$, $k$ is the Boltzmann constant, $E_u$ is the energy of the upper state, $T_\mathrm{ex}$ is the excitation temperature, and $Q_\mathrm{rot}$ is the rotational partition function which we calculate by summing over each of the energy levels. For this calculation of $Q_\mathrm{rot}$, we take the energy levels and statistical weights of each state from the CDMS catalog \citep{CDMS}. We use $E_u=\SI{5.75}{K}$ for the HC$_5$N $J=9\shortrightarrow8$ transition and $E_u=\SI{1.606}{K}$ for the CCS $J_N=2_1\shortrightarrow1_0$ transition, which are taken from \cite{1976JMoSp..62..175A} and \cite{1987ApJ...317L.115S} respectively.
 
To calculate a column density of \ce{HC7N} using the stacked spectrum, we use equations 1 and 2. For \ce{HC7N}, we use $B=\SI{564}{MHz}$, $\mu=\SI{5.0}{D}$, and $E_u= \SI{11.93}{K}$ which is the average of the upper energies of the $J=20\shortrightarrow19$ and $J=21\shortrightarrow20$ transitions, since the stacked spectrum is an average of both transitions from all sources. We compute the partition function by summing over the energy states of \ce{HC7N} like above. In order to compute a ratio of \ce{HC5N} to \ce{HC7N} we also generate a stacked spectrum of \ce{HC5N}, which is shown in Figure \ref{fig:HC5N_stack}, and use it to calculate a  column density of stacked spectra of \ce{HC5N}. We then use the stacked column density of \ce{HC7N} and \ce{HC5N} to calculate a ratio of \ce{HC5N} to \ce{HC7N}.

We calculate the column densities of CCS using the same method for the rotational partition function and relation of upper energy column density to total column density, but the equation for the upper state column density ($N_u$) that was used for CCS is given as 
\begin{equation}
    \frac{N_u}{g_u}=\frac{3k\int T_Bdv}{8\pi^3\nu\mu^2S}
\end{equation}
where $\nu$ is the frequency of the transition and $S$ is the intrinsic line strength. We use an intrinsic line strength of 1.98 for the CCS $J_N=2_1\shortrightarrow1_0$ transition which is taken from \cite{Suzuki1992}. For the rotational constant of CCS we use 6477.75 MHz and for the dipole moment we use \SI{2.88}{D}, both of which were taken from \cite{1987ApJ...317L.115S}. 

Due to the comparatively low gas density ($\sim\!\SI{e4}{cm^{-3}}$) we cannot necessarily assume that the level populations are in LTE described by $T_{ex}=T_{kin}$ when calculating column densities for these species. With only one line measured for each of the carbon chain species here, we are not able to constrain the excitation temperature of the gas directly. \cite{pillai2006} measured the excitation temperature of \ce{NH3} in 20 infrared-dark clouds and found a median excitation temperature of \SI{5.7}{K}. Similarly, \cite{Vastel18} measured a sub-thermal excitation temperature in the range of $4.9-\SI{5.6}{K}$ for \ce{CCS} in the pre-stellar core L1544. \citep{Kalenski04} measured the rotational temperature of \ce{HC5N} and CCS in the TMC-1 dark cloud to be 4.3 and 5.7 K respectively and \cite{Bianchi23} measured the excitation temperature of \ce{HC5N}, \ce{HC7N}, and \ce{HC9N} in L1544 and found that they are moderately sub-thermally populated with an excitation temperature in the range 6 to 7.5 K. The carbon chains in L1544, however, from in a region of low gas density (100 cm$^{-3}$) that has exposure to the interstellar radiation field \citep{Bianchi23}, so their excitation temperature may not best represent our sources compared to the other sources mentioned above. Based on these observations, we adopt an excitation temperature range of 4-6 K for our column density calculations, with a central temperature of 5 K used for the reported column densities. The range of excitation temperatures ($4-\SI{6}{K}$) we use is propagated as a source of uncertainty in the column density calculation.

\subsection{Calculation of Clump Densities}
To calculate the number density of our sources, we use data from the APEX Telescope Large Survey of the Galaxy (ATLASGAL), at \SI{870}{\um} \citep{Schuller,ATLASGAL}. This data is used rather than the Bolocam Galactic Plane Survey \SI{1.1}{mm} data due to higher SNR and superior angular resolution (\ang{;;19} versus \ang{;;33}). Using the ATLASGAL continuum images of our 11 sources with carbon chain detections, we measure the \SI{870}{\um} flux using the same \ce{NH3} mask that was used to extract the carbon chain spectra. This is done to ensure that the clump \ce{H2} column densities and number densities are calculated for the same spatial region as the carbon chain column densities. This enables us to calculate the abundances of the carbon chain molecules relative to \ce{H2} over the same spatial region. As shown in Figure \ref{fig:ATLASGAL_nh3}, the \ce{NH3} mask appears to trace the bulk continuum emission in the ATLASGAL images, and thus the region where the carbon chain spectra is extracted also traces the bulk of the dust continuum emission. Note, that although the \ce{HC5N} and \ce{NH3} coincide in the 2D images, there is still the possibility that there is stratification of the molecules in these clumps, and the \ce{HC5N} and \ce{NH3} may not arise from the same spatial location in the clump.

For G24051, one of the sources with the brightest carbon chain detection, the \ce{HC5N} emission appears to be within the \ce{NH3} emitting region (see Figure \ref{fig:b}) which is within the 2D region of the bulk dust continuum emission. The rest of the moment 0 maps of \ce{HC5N} for the other sources are shown in the Appendix. The detection of carbon chains in 11 of our sources using the \ce{NH3} mask does suggest that the carbon chains and \ce{NH3} do share similar spatial locations. If, however, the bulk of the carbon chains is not entirely cospatial with the \ce{NH3} and thus the dust, then we would underestimate the carbon chain fluxes and thus their abundances relative to \ce{H2} This would ultimately cause an overestimate in the age estimation, so our conclusions about the evolutionary stage of the sources are not changed. Regardless, the carbon chain and \ce{H2} column densities are calculated over the same spatial region corresponding to the area of the bulk dust emission, so the abundance of the carbon chain molecules is appropriate for the shown spatial regions. After measuring the \SI{870}{\um} flux from the ATLASGAL images, we calculate the gas mass using the equation 
\begin{equation}
    M_\mathrm{gas}=\frac{d^2F_{\nu}R}{B_{\nu}(T)\kappa_{\nu}}
\end{equation}
where $d$ is the distance to the sources, $F_{\nu}$ is the integrated flux, $R$ is the gas-to-dust ratio, $B_{\nu}(T)$ is the Planck Function, and $\kappa_{\nu}$ is the dust opacity \citep{Schuller}. We use the dust temperatures derived from \ce{NH3} on \ang{;;30} scales reported in \cite{svoboda19}. We use a dust opacity of $\kappa_{\nu}=\SI{1.85}{cm^2.g^{-1}}$ for dust grains with thin ice mantles coagulated at densities of \SI{1e5}{cm^{-3}} for \SI{1e5}{yr} years \citep{Opacity}. We then calculate the \ce{H2} mass of each clump by assuming a gas-to-dust ratio of 100. 

We convert the gas masses that we calculate into the number of \ce{H2} molecules and then divide by the volume of the clump to calculate the number density of \ce{H2}. We estimate the volume of the clumps by first finding the cross sectional area of each of the clumps from the masking aperture, and then approximating this area as the cross-sectional area of a sphere and solving for a radius ($R=\sqrt{A/\pi}$). We then use the radius of each source to calculate the volume of each clump. We calculate the \ce{H2} column density by dividing the the total number of \ce{H2} molecules by the area of the extraction aperture. These column densities are then used to compute the abundance of the carbon chain molecules to \ce{H2} for comparison with the chemical models.

\subsection{Trends in Column Densities}
Correlations in column densities of molecules do not directly mean that there is a chemical link between the two molecules (e.g. \citealt{belloche20}). Instead, it may suggest a common precursor or that the two species are formed through similar physical processes \citep{Suzuki1992,Quenard18}. We first look for trends in the column densities of the carbon chain molecules and compare the column densities with clump kinetic temperature. The kinetic temperatures are derived from the \ce{NH3} observations and will presented in an upcoming manuscript (Svoboda et al. in prep.). They were calculated using the cold \ce{NH3} approximation in \cite{rosolowsky08} and fitting with PySpecKit \citep{Pyspec}. Scatter plots of HC$_5$N column density and temperature and CCS column density and HC$_5$N column density can be seen in Figure \ref{fig:g}. We find a positive linear relationship between CCS column density and HC$_5$N column density. We calculate a Pearson's $r$ value of 0.70 for the relationship between the \ce{HC5N} and CCS column densities. This gives a 95$\%$ confidence interval of [0.18,0.92] for the Pearson's $r$ coefficient. This correlation between CCS and \ce{HC5N} column densities could indicate that these molecules form from similar processes \citep{Quenard18}. We do not find a significant correlation between the \ce{HC5N} column density and the clump temperature. 

\subsection{Comparison with Chemical Models}

In order to compare the carbon chain abundances to chemical models, we also calculate their abundances relative to H$_2$. We calculate the abundances by converting the \ce{H2} mass of the clumps into a column density by dividing by the area of the clump and then dividing our measured carbon chain column densities by the \ce{H2} column densities. The calculation of the \ce{H2} mass of these SMDCs is described above. Uncertainty in the abundances are calculated by propagating the error of clump mass, excitation temperature, and column density using general error propagation formulae. We also include the covariance between the Gaussian width and amplitude in the uncertainty calculations. The results of the column density and abundance calculations can be seen in Table \ref{table:b}. 

We compare our calculated abundances of HC$_5$N and CCS to the UMIST 13 Dark Cloud Chemistry Models, which describes the abundance of molecules over the age of a dark molecular cloud \citep{2013A&A...550A..36M}. Although grain surface chemistry does play a significant role in the formation of various molecules, the Dark Cloud UMIST 13 model is limited to gas phase reactions. This is primarily due to the many uncertainties in grain surface chemistry that prevent quantitatively accurate predictions \citep{2013A&A...550A..36M}. 

In comparing our measured abundances to the models, we find that our measured abundances for CCS in all of the 11 sources with detections are not in agreement with the UMIST chemical models. Our measured abundance values are consistently at least an order of magnitude greater than what the UMIST models predict at all evolutionary timescales. The CCS abundance predicted by the UMIST model peaks at the same time as the \ce{HC5N} abundance ($\sim$10$^5$ years) and has a peak abundance of 1$\times10^{-11}$. The median abundance of CCS measure for our sources is 1$\times10^{-10}$, an order of magnitude higher than what UMIST predicts. It is noted by the authors of the UMIST 13 model that their model does not agree with previous measurements of CCS in infrared dark clouds, so this disagreement is a well known issue. This is likely due to the large rate coefficient for the destruction of \ce{CCS} with atomic oxygen \citep{2013A&A...550A..36M}. This specific rate coefficient is lacking low temperature measurements and is assumed to be temperature independent \citep{Loison12,2013A&A...550A..36M}. This has a large impact on the calculated CCS abundances, causing the predicted abundances of CCS from the UMIST dark cloud model to be low \citep{2013A&A...550A..36M}. Because of this disagreement with the UMIST models, we do not use CCS to estimate the chemical evolutionary ages of these sources. 

The measured abundance values of \ce{HC5N} from our sources, however, are comparable to the UMIST 13 dark cloud chemistry model. We run three UMIST chemical models, all of which have all the same input parameters except the clump density. The input parameters that we use for the models are $T_k=\SI{12}{K}$, $A_v=10$, cosmic ionization rate equal to the local Galactic disk ($\zeta_0=1.3\times10^{-17}$ s$^{-1}$), UV radiation field equal to local Galactic disk from \cite{draine78}, and initial abundances equal to values atomic (see Table 3 of \cite{2013A&A...550A..36M} and our Table 4). From our calculations of clump densities, we find that the median density of our sources is $n$(H$_2$)=2$\times10^4$ cm$^{-3}$, which is the main density we consider for the analysis. We calculate models for three different clump densities,  $n(\ce{H2})=\SI{1.2e4}{cm^{-3}}$, $n(\ce{H2})=\SI{2.0e4}{cm^{-3}}$, and $n(\ce{H2})=\SI{3.1e4}{cm^{-3}}$. The other two densities we use are the median density plus/minus the median absolute deviation of the densities for these sources and are used to illustrate another source of uncertainty in determining the chemical ages of these sources.

\begin{table}[!htbp]
    
    \centering
    
    \caption{Initial abundances relative to total H nuclei ($n_H$).}
    
    \begin{tabular}{c c c c }
    
    \hline
    
        Species i & $n_i/n_H $&Species i & $n_i/n_H $ \\
        \hline
       \ce{H2} &0.5 &Na & 2.0$\times10^{-9}$  \\
        H&5$\times10^{-5}$ &Mg &7.0$\times10^{-9}$  \\
       He & 0.09 & Si & 8$\times10^{-9}$   \\
       C &1.4$\times10^{-4}$ &P &3.0$\times10^{-9}$   \\
       N &7.5$\times10^{-5}$ &S & 8.0$\times10^{-8}$  \\
       O & 3.2$\times10^{-4}$ &Cl & 4.0$\times10^{-9}$  \\
       F &2.0$\times10^{-8}$ &Fe & 3$\times10^{-9}$  \\
         
         \hline
    \end{tabular}
    \begin{minipage}{6 cm}
    \vspace{0.3cm}
    
     \textbf{Note.} Taken from \cite{2013A&A...550A..36M}. 
    \end{minipage}
    \label{tab:inital_abund}
\end{table}

The result of comparing our measured HC$_5$N abundances to \ce{H2} for the 11 sources with detections and the dark cloud chemistry models can be seen in Figure \ref{fig:h}. Comparing with the $n(\ce{H2})=\SI{2e4}{cm^{-3}}$ UMIST model, we find that it favors that all of the clumps are less than approximately \SI{0.5}{Myr} old at this density. We determine this maximum age limit by finding the largest time at which our measured abundances intersect the abundances predicted by the dark cloud chemistry models. There are additional uncertainties in the modeling carbon chain chemistry which are described in section 5.2. This calculation of the absolute abundances of \ce{HC5N} to \ce{H2} using the \ce{NH3} extraction aperture does not take into consideration potential stratification of the clump.

We also compare our value of the ratio of \ce{HC5N} to \ce{HC7N} to the ratio predicted by the UMIST 13 dark cloud chemistry models. This is shown in Figure \ref{fig:HC7N_ratio}. By determining where our measured \ce{HC5N} to \ce{HC7N} ratio intersects with the ratio predicted by the $n(\ce{H2})=\SI{2e4}{cm^{-3}}$ UMIST model, we find that our sources are less than approximately \SI{1}{Myr} old, which is consistent with our age estimation from the \ce{HC5N} abundances relative to \ce{H2}.

\begin{table*}[!ht]
 
\centering
\caption{Values for column density and abundance of HC$_5$N and CCS for each source with a detection. }
\begin{tabular}{c c c c c c }
    \hline
   
    Source & Molecule& Column Density (cm$^{-2}$ $*10^{12}$) & \ce{H2} Column Density (cm$^{-2}$ $*10^{22}$) &\ce{H2} Density (cm$^{-3}$ $*10^{4}$) & Abundance to \ce{H2} ($*10^{-11}$)  \\
      
    \hline
        G22695 & HC$_5$N &1.4$^{+0.4}_{-0.3}$&4.0$\pm$0.5&3.1$\pm$0.5& 3.5$\pm$0.8 \\
        G23297 & HC$_5$N &3.0$^{+0.5}_{-0.3}$&3.3$\pm$0.4 &2.3$\pm$0.4 & 10$\pm1$\\
         G23481 & HC$_5$N&1.5$^{+0.4}_{-0.3}$ & 4.2$\pm$0.8 &2.0$\pm$0.8 & 3.6 $\pm$0.9 \\
         G24051 & HC$_5$N&4.2$^{+0.6}_{-0.3}$ & 4.0$\pm$0.4 &2.8$\pm$0.4 &11$\pm$1  \\
         G28539& HC$_5$N&1.6$^{+0.2}_{-0.2}$ & 3.2$\pm$0.4 & 1.3 $\pm$0.4 &5.0$\pm$0.7 \\
         G28565& HC$_5$N& 5.8$^{+0.9}_{-0.6}$&6.1$\pm$0.9 &5.0$\pm$0.9&10$\pm$2  \\
         G29558& HC$_5$N &3.5$^{+0.5}_{-0.3}$ &2.9$\pm$0.5 &1.4$\pm$0.5 &12$\pm$2 \\
         G29601 & HC$_5$N&1.0$^{+0.3}_{-0.3}$ &0.9$\pm$0.3 & 1.0$\pm$0.6&11$\pm$3 \\
         G30120 & HC$_5$N&0.96$^{+0.2}_{-0.1}$ &2.4$\pm$0.4 &1.4$\pm$0.4 &4.0$\pm$0.8 \\
         G30660 & HC$_5$N&1.6$^{+0.2}_{-0.1}$ &4.0$\pm$0.4 &2.6$\pm$0.4 & 4.0$\pm$0.5\\
         G30912 & HC$_5$N& 5.8$^{+0.8}_{-0.7}$ & 6.1$\pm$0.7& 6.5$\pm$0.7 &9.5$\pm$0.9 \\
         G22695 & CCS &3.0$^{+0.8}_{-0.6}$ & 4.0$\pm$0.5 &3.1$\pm$0.5 &8$\pm$2 \\
        G23297 & CCS  &8$^{+2}_{-1}$ &3.3$\pm$0.4 &2.3$\pm$0.4 &24$\pm$6 \\
         G23481 & CCS &4.5$^{+0.9}_{-0.7}$ &4.2$\pm$0.8 &2.0$\pm$0.8 &11$\pm$2 \\
         G24051 & CCS & 4.9$^{+0.9}_{-0.8}$&4.0$\pm$0.4 &2.8$\pm$0.4& 12$\pm$2 \\
         G28539& CCS & 5.2$^{+0.9}_{-0.8}$ &3.2$\pm$0.4 &1.3$\pm$0.4 & 16$\pm$3 \\
         G28565& CCS &15$^{+3}_{-2}$ &6.1$\pm$0.9 &5.0$\pm$0.9 & 25$\pm$6\\
         G29558& CCS  & 8$^{+2}_{-1}$ &2.9$\pm$0.5 &1.4$\pm$0.5 &28$\pm$5 \\
         G29601 & CCS &  3.7$^{+0.8}_{-0.6}$&0.9$\pm$0.5 &1.0$\pm$0.6 &41$\pm$10 \\
         G30120 & CCS &  5.4$^{+0.9}_{-0.8}$&2.4$\pm$0.4 &1.4$\pm$0.4 &23$\pm$5 \\
         G30660 & CCS & 3.6$^{+0.7}_{-0.4}$&4.0$\pm$0.4 &2.6$\pm$0.4 &9$\pm$2 \\
         G30912 & CCS &  6.1$^{+0.9}_{-0.9}$&6.1$\pm$0.7 &6.5$\pm$0.7 &10$\pm$2 \\
          Stack & \ce{HC7N} &  1.2$^{+0.8}_{-0.7}$&N/A& N/A&N/A \\
      \hline
    \end{tabular}
\label{table:b}
 \begin{minipage}{17cm}
    \vspace{0.3cm}
    
     \textbf{Note.} The column densities of the carbon chains are calculated with an excitation temperature of 5 K, using a range of excitation temperatures from 4-6 K for the uncertainties. The \ce{H2} column densities are computed over the same spatial areas as the carbon chain spectra from the ATLASGAL \si{870}{\micron} images of these sources and the abundances of the carbon chains relative to \ce{H2} are calculated by dividing the column densities.
    \end{minipage}

\end{table*}

\begin{figure*}[h!]
\centering
\includegraphics[scale=0.9]{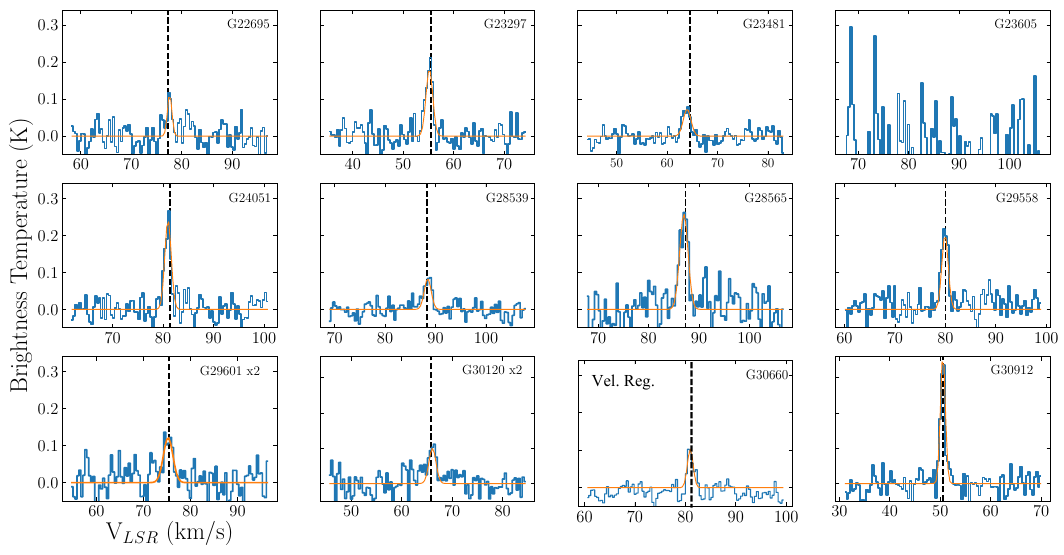}
\caption{Spectra of HC$_5$N for all 12 of our sources (blue) with best fit Gaussians (yellow). The only non-detection of HC$_5$N in on 12 sources occurs in G23605, which is the upper right hand corner spectrum in this figure. The black dashed line shows the systematic velocity from \cite{svoboda16}. G30660 is the only spectrum shown acquired through the velocity registration method. G29601 and G30120 are scaled for clarity. }\label{fig:d}
\end{figure*}
\begin{figure*}[h!]
\centering
\includegraphics[scale=.9]{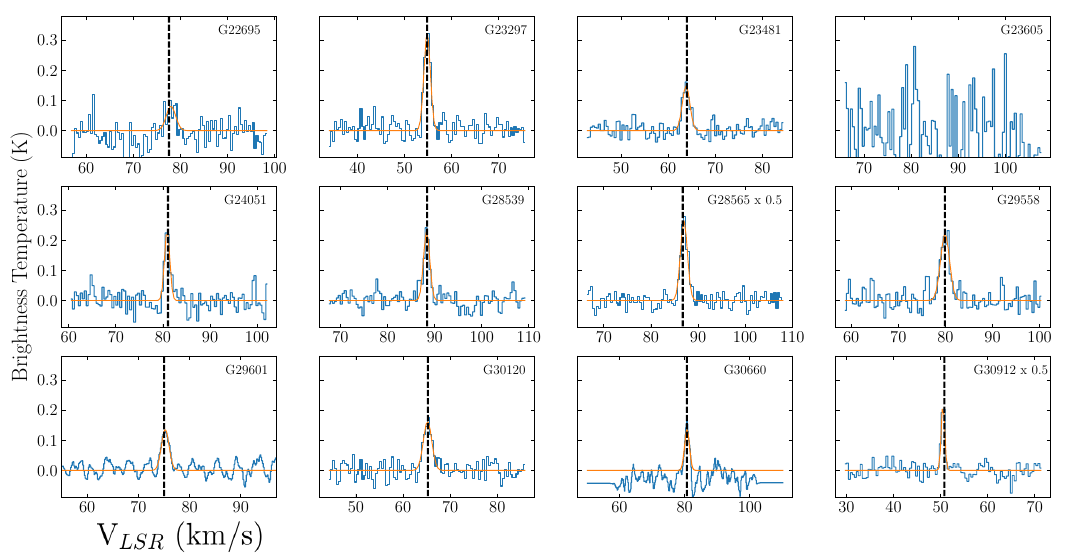}
\caption{Spectra of CCS for all 12 of our sources (blue) with best fit Gaussians (yellow). The only non-detection of CCS in on 12 sources occurs in G23605, which is the upper right hand corner spectrum in this figure. The black dashed lines show the systematic velocity. Velocity registration was used to obtain the spectra of G30660 and G29601.}\label{fig:e}
\end{figure*}

\begin{figure}[!htbp]
\includegraphics[scale=.65]{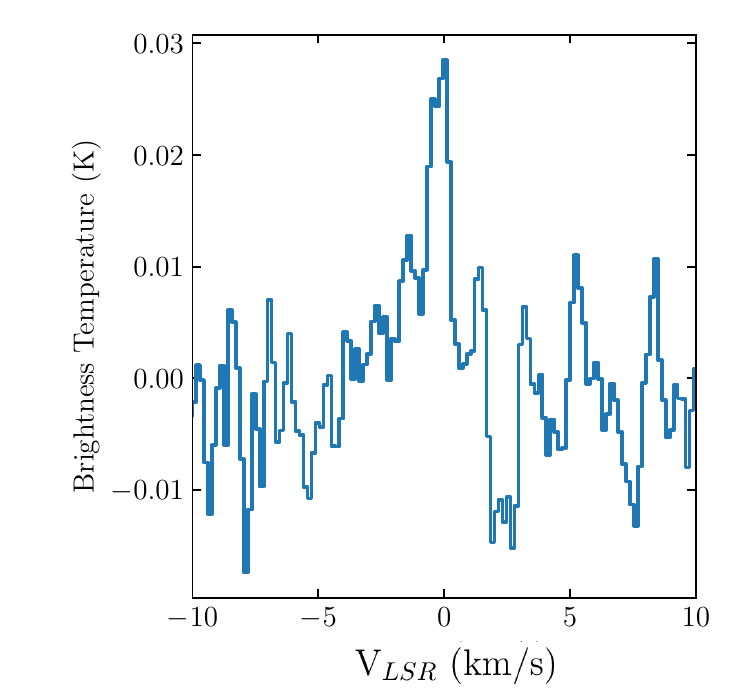}
\caption{Stacked spectrum of HC$_7$N. The stack includes all observed transitions of \ce{HC7N} for all sources with \ce{HC5N} detections.  }\label{fig:f}
\end{figure}

\begin{figure}[!htbp]
\includegraphics[scale=.65]{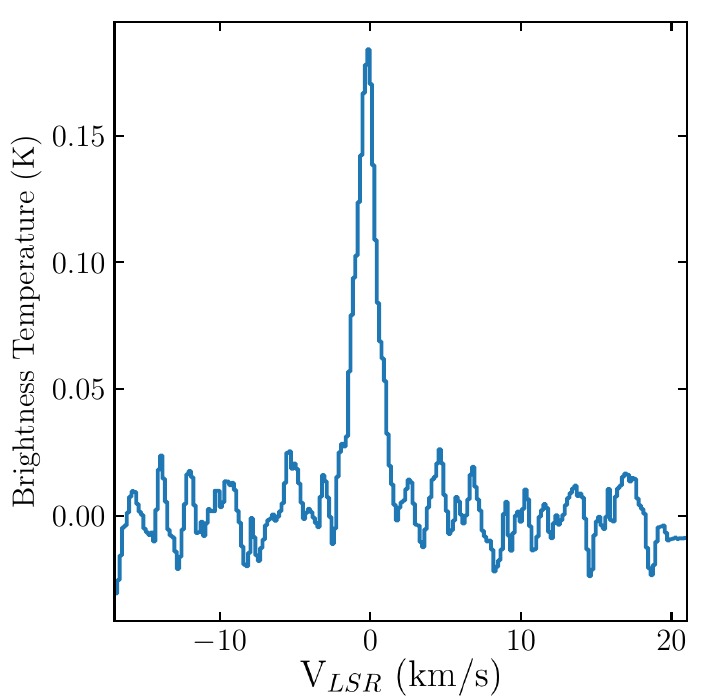}
\caption{Stacked spectrum of HC$_5$N. The stack includes all sources of \ce{HC5N} with detections.  }\label{fig:HC5N_stack}
\end{figure}

\begin{figure*}[!htbp]
\centering
\includegraphics[scale=.63]{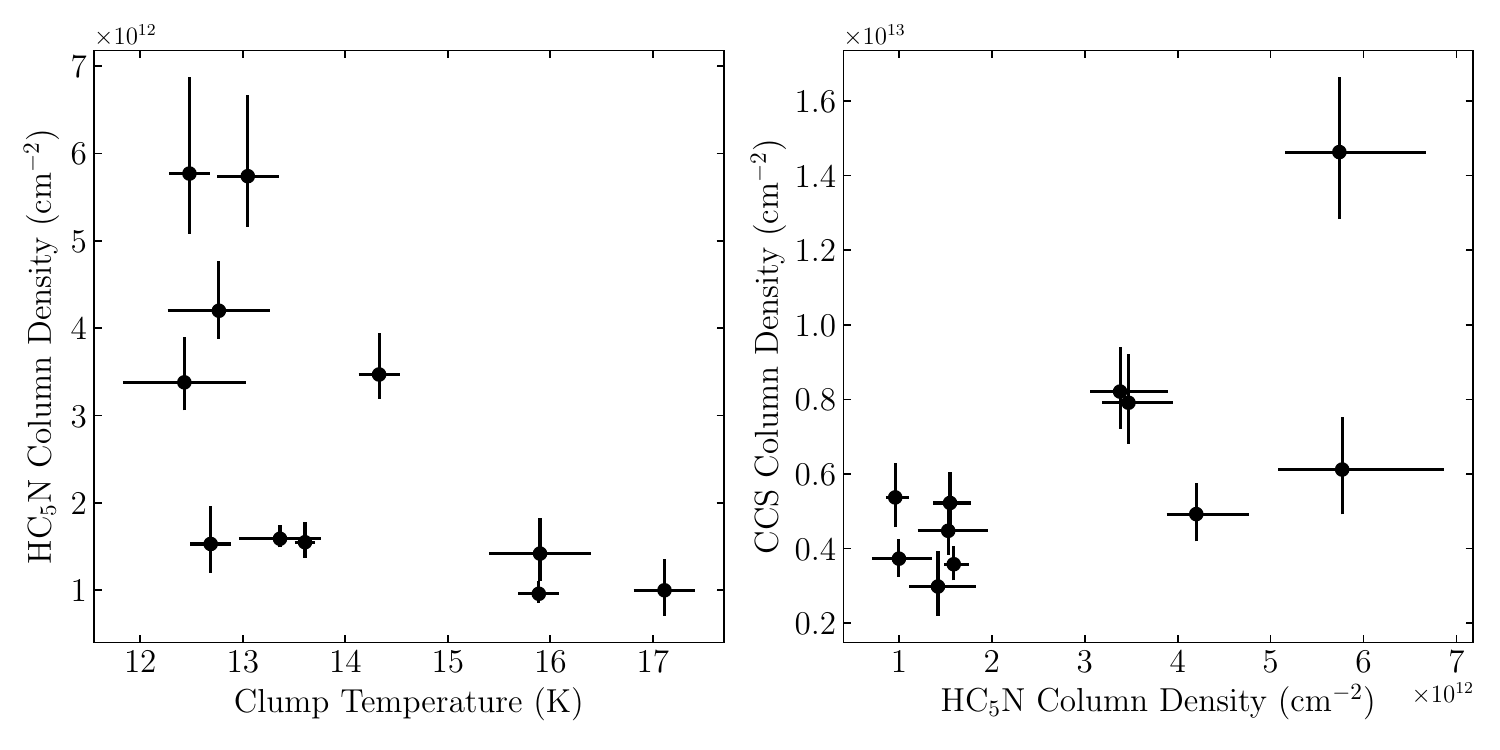}
\caption{Plot on left shows HC$_5$N column density as a function of the clump kinetic temperature.  Figure on the right shows CCS column density as a function of HC$_5$N column density. }\label{fig:g}
\end{figure*}

\begin{figure*}[!htbp]
\centering
\includegraphics[scale=.5]{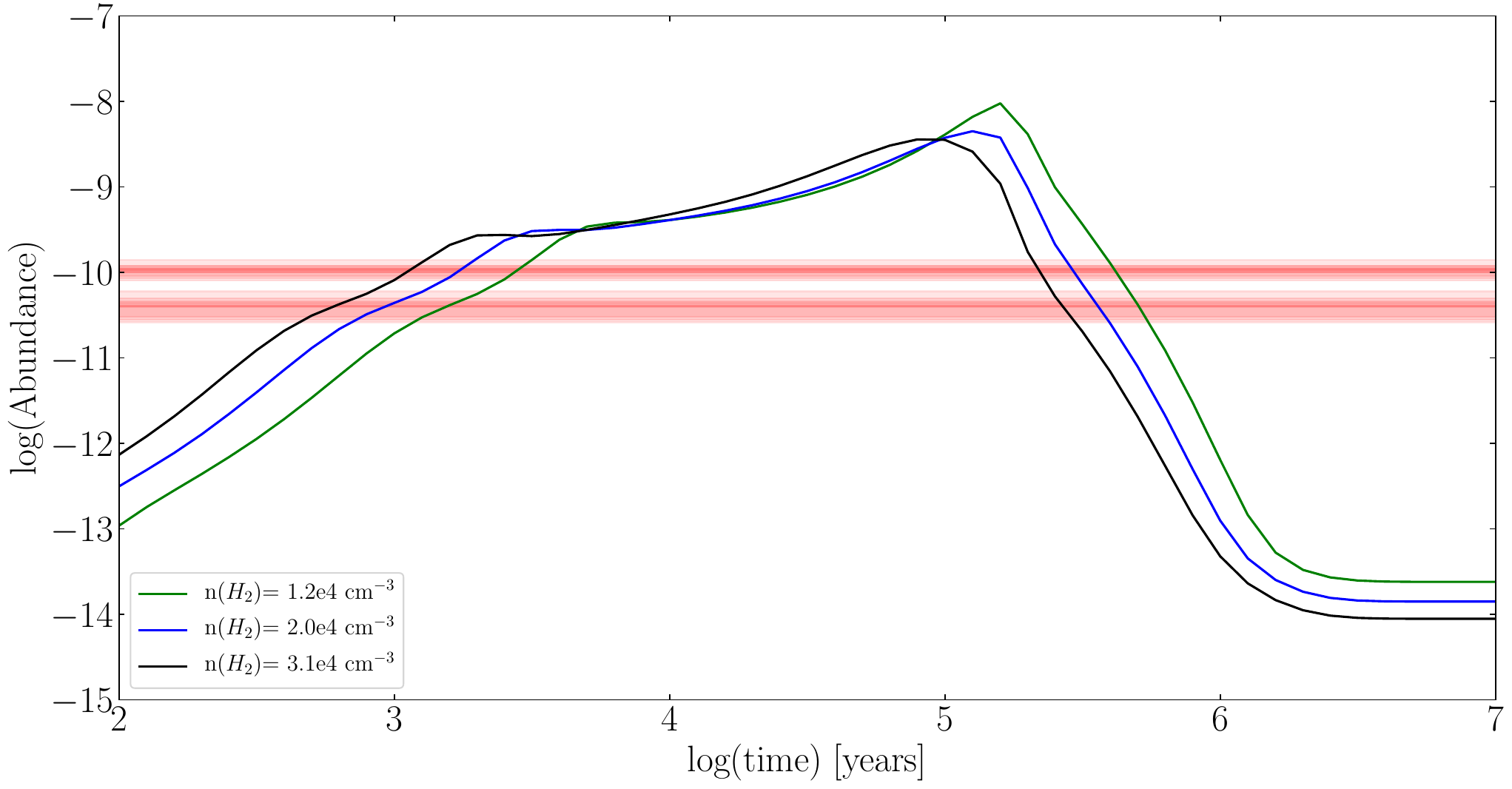}
\caption{Plot of abundance of HC$_5$N as a function of age of the clump. The blue line shows the \ce{HC5N} abundance predicted by the UMIST 13 chemistry model using a clump density of  $n$(H$_2$)=2.0$\times$10$^4$ cm$^{-3}$, the black line shows a model with a density of $n$(H$_2$)=3.1$\times$10$^4$ cm$^{-3}$, and the green line shows a model with a density of $n$(H$_2$)=1.2$\times$10$^4$ cm$^{-3}$. The red colored areas represent the uncertainty in the measured \ce{HC5N} abundances of each of our 11 sources.}\label{fig:h}
\end{figure*}
\begin{figure}
    \centering
    \includegraphics[scale=.6]{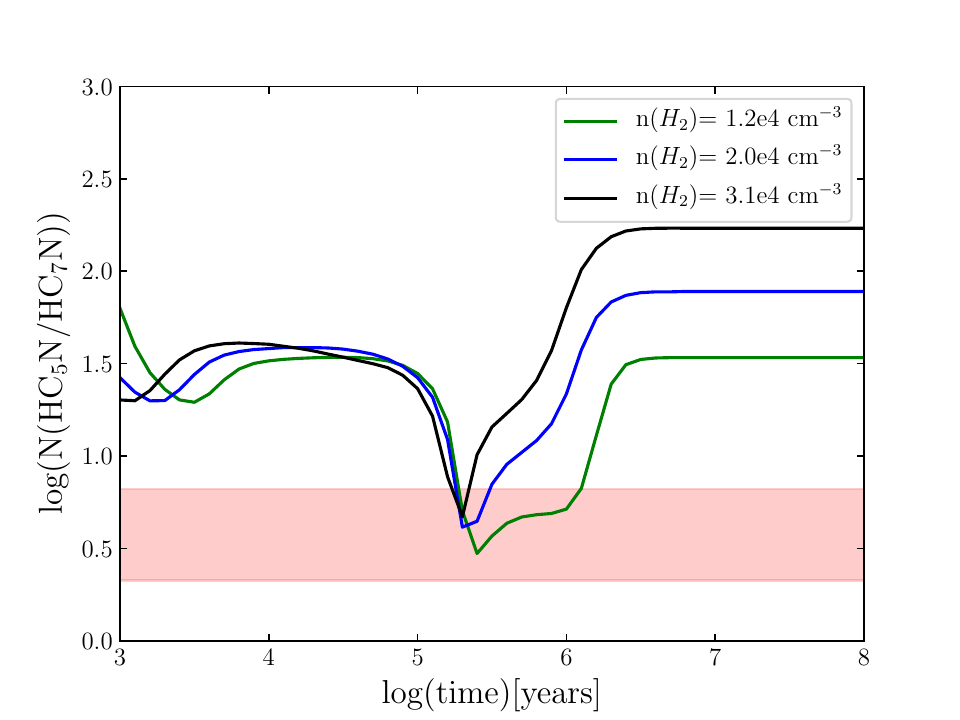}
    \caption{Ratio of \ce{HC5N} to \ce{HC7N} as a function of time. The blue, green, and black lines show the abundances predicted by the UMIST 13 dark cloud chemistry model at the densities indicated and the red colored area shows our measured ratio with uncertainties. }
    \label{fig:HC7N_ratio}
\end{figure}

\section{Discussion}\label{sec:Discussion}
The potential for sources like these SMDCs to form high-mass stars is an open question in HMSF. Are quiescent clouds like these actually the site of HMSF and are they useful for studying the initial conditions of HMSF? Here we discuss the implications in regards to HMSF of the ages we estimate for these SMDCs as well as the uncertainties involved in this method of age estimation using observations of carbon chain molecules.

\subsection{Reactions Dominating Carbon Chain Production and Destruction}
When \ce{HC5N} is at its peak abundance models at \num{2e5} years, the UMIST models predict that the production of \ce{HC5N} is dominated by two reactions: 

\begin{equation}
    \ce{N}+\ce{C5H2}\shortrightarrow \ce{HC5N}+\ce{H}
\end{equation}
and
\begin{equation}
    \ce{HC5NH+}+\ce{e-}\shortrightarrow \ce{HC5N}+\ce{H}
\end{equation}
where the dominant reaction that produces \ce{HC5NH+} is
\begin{equation}
    \ce{CH4}+\ce{C4N+}\shortrightarrow\ce{HC5NH+}+\ce{H2}.
\end{equation}

Similarly, there are two reactions that dominate the destruction of \ce{HC5N} when it is at its peak abundance:
\begin{equation}
    \ce{C+}+\ce{HC5N}\shortrightarrow \ce{C5H+}+\ce{CN}
\end{equation}
and
\begin{equation}
    \ce{HCO+}+\ce{HC5N}\shortrightarrow \ce{HC5N+}+\ce{CO}.
\end{equation}
 These reactions are found by analyzing the formation and destruction rates in the UMIST 13 models at the time when \ce{HC5N} is at its maximum abundance. \cite{Taniguchi17} and \cite{Burkhardt} suggested that the main formation mechanism for \ce{HC5N} is the reactions between nitrogen atoms and hydrocarbons, like shown by the UMIST model. There are, however, other reactions that could contribute to the formation the carbon chain molecules \citep{Bianchi23} (see below) and the contributions of these different reactions may not be accurately accounted for in the UMIST chemical model.

\subsection{Uncertainties in Chemistry Models}

A possible limitation to this method of determining ages of the molecular clumps are the uncertainties involved in the dark cloud chemistry models. To determine how much of an effect the model uncertainties have on our age estimation, we test how different input temperatures and densities of the models affect our  measured ages for these SMDCs. We test a range of temperatures from $10-\SI{40}{K}$ and a range of densities from $n$(\ce{H2})=\num{1e3} cm$^{-3}$ to $n$(\ce{H2})=\num{1e5} cm$^{-3}$, which is a typical range of densities for molecular clumps \citep{Rathborne2006,Clumps} and covers the temperature and density range of our 12 sources \citep{svoboda19}.

\subsubsection{Temperature}
In testing the input temperature, we find that a higher input temperature corresponds to a higher peak abundance of \ce{HC5N}, but does not change the time at which our measured \ce{HC5N} abundances intersect with the UMIST model. The input density of the UMIST models, however, does change the age that we estimate for these clumps. We find that a higher density corresponds to the peak \ce{HC5N} abundance occurring at earlier times and a lower density corresponds to the peak \ce{HC5N} abundance occurring at later times. Even at the lowest measured clump density of 
$n$(\ce{H2})=\num{1e3} cm$^{-3}$ the chemical ages estimated for these clumps are still $\lesssim$ 1 Myr.
\subsubsection{UV Radiation Field}
We also test the UV radiation field and extinction ($A_V$) as input parameters for the chemical models. All sources lie within the galactic plane ($|b|<1.5^{\circ}$) at similar Galactocentric radii \citep{svoboda19}, so we don't expect to have a UV radiation field less than the standard Galactic disk for any source. Therefore, we only test values greater than the UV radiation field of the Galactic disk. We test values of the UV radiation field up to 5 times the Galactic disk radiation field. We test the different values of UV radiation field along with different extinctions. We estimate the ATLASGAL beam average $A_v$ values along the line of sight between us and the clumps for our 12 sources using the equation 
\begin{equation}
    N(\ce{H2})=1\times10^{21} A_v \hspace{1mm} \si{cm^{-2}}
\end{equation}  
where $N(\ce{H2})$ is the \ce{H2 column density} \citep{Yamamoto17}. The minimum \ce{H2} column density from our sources (see Table 5) corresponds to an extinction value of $A_v=9$ mag. This extinction is calculated over the entire ATLASGAL beam and is the extinction on out line of sight to these clumps. Because of this, we run chemical models down to an extinction value of $A_v=5$ mag to test the effect of extinction of the carbon chain abundance.   

We find that even with a UV radiation field that is 5 times stronger than the standard Galactic disk assumed in the UMIST models, and an extinction value in the range $A_V=5-10$ mag, the UMIST models still predict \ce{HC5N} abundances that do not reach 10$^{-12}$ at times later than 1 Myr. We find that increasing the UV radiation field and decreasing the extinction makes the \ce{HC5N abundance} reach its peak at earlier times and gives a larger \ce{HC5N} abundance at times greater than 1 Myr. However, the \ce{HC5N} abundances predicted at 1 Myr and longer is still below all of our abundance estimations by at least an order of magnitude. Therefore, the chemical age of $\lesssim$ 1 Myr appears to be robust against the of level exposure to UV radiation. We also test extinction values that range from $A_V=10-15$ mag. We find that for extinction values between $A_V=10-15$ mag, the difference in the model predicted carbon chain abundance is negligible (less than 5$\%$) and the age estimation of the clumps is unchanged.

\subsubsection{Chemical Network}
Our choice of chemical model has the potential to change the estimation of the age of these clumps. \cite{Taniguchi2019}, using the gas-grain code Nautilus \citep{Ruaud16}, also predict the abundance of \ce{HC5N} and \ce{HC7N} as a function of time within starless clouds. Their models also indicate that these carbon chains are an early time species in the lifetime of starless clouds and their abundances of both \ce{HC5N} and \ce{HC7N} drop below \num{1e-16} after 0.5 Myr, similar to the predictions made by the UMIST models used in this analysis. This consistency between the two sets of models provide further support that these carbon chain species are significantly depleted at times greater than 1 Myr and the abundances on the order of \num{1e-11} we measured for \ce{HC5N} in our sources indicate that these clumps are at an early evolutionary stage.

There are also uncertainties in the underlying chemistry of the carbon chain molecules in the UMIST model. The reactions that dominate the production and destruction of \ce{HC5N} in the UMIST model are listed in section 5.1. However, there are other gas phase reactions that are thought to contribute to the formation of carbon chain molecules like \ce{HC5N}. Some of these other reactions are \ce{C4H2}+\ce{CN}$\shortrightarrow$ \ce{HC5N}+\ce{H},  \ce{C6HN}$\shortrightarrow$ \ce{HC5N}+\ce{C}, and  \ce{H2C5N+}+e$^+$$\shortrightarrow$ \ce{HC5N}+\ce{H} \citep{Fukuzawu98,Bianchi23}. The uncertainty in the rate constants of the reactions creating and destroying the carbon chain molecules can lead to large uncertainties in the predicted abundances. For a further review on the lack of rate constants for the reactions listed above, see \cite{Bianchi23}.

For simple species, \cite{Vasyunin04} found that the uncertainty in the rates in the UMIST chemical network can lead to an order of magnitude of uncertainty on the predicted abundances of the chemical models. However, this uncertainty could be larger for more complex molecules. Furthermore, \cite{Loison14} found that abundance of \ce{HC5N} is dependent on the assumed initial C/O ratio. Because of the uncertainties in the chemical network and initial abundances, we expect that there is at least an order of magnitude uncertainty on the carbon chain abundances predicted by the UMIST models. Our \ce{HC5N} abundances for these 
11 sources, however, are 3-4 orders of magnitude greater than the prediction of the UMIST models for times greater than 1 Myr. These uncertainties of 1-2 orders of magnitude in the chemical model likely do not change the result that the detection of carbon chain molecules in these clumps indicates that they are less than 1 Myr old at these densities.

\subsubsection{\ce{H2 Column Density}}
The \ce{H2} column density contributes to the uncertainty in the carbon chain abundance and is thus a source of uncertainty in our age determination. We calculate the \ce{H2} column density towards these sources from ATLASGAL data, which has an angular resolution of 19'' and a field of view of $5\times5$ arcminutes \citep{schuller09}. The field of view of our VLA data-cubes are $\sim\!3\times3$ arcminutes, and thus the entirety of the clumps viewed in VLA data are within the field of view of the ATLASGAL \SI{870}{\micron} images. By using the exact same spatial extraction aperture for ATLASGAL images and the VLA carbon chain data cubes, we ensure that the measurement of the \ce{H2} column density is from the same spatial region as the carbon chain spectra, allowing for a direct calculation of the carbon chain abundance relative to \ce{H2} in that spatial region. Since there is not significant \ce{NH3} or dust continuum emission outside of the \ce{NH3} extraction apertures, then using the \ce{NH3} masks as extraction apertures for the carbon chain spectra could result in an underestimate of the carbon chain line flux in these sources if the carbon chains have a different spatial distribution than the \ce{NH3}. This would then lead to an underestimation of the carbon chain abundances and an overestimation of the ages of the clumps. Therefore, our conclusion that these clumps are young (less than 1 Myr) is largely unaffected by our choice of extraction aperture. The moment 0 maps of \ce{HC5N} shown in the Appendix demonstrate that there is no significant sources of \ce{HC5N} emission outside of the \ce{NH3} contours in any source, so we are likely not missing significant \ce{HC5N} flux. The sources with \ce{HC5N} detected in their moment 0 maps have their \ce{HC5N} emission within the \ce{NH3} contours as shown in Figure \ref{fig:hc5n_mom_maps}. This indicates that the choice of using \ce{NH3} as a carbon chain extraction aperture does not have a significant effect on our results. Furthermore, the \ce{HC5N} to \ce{HC7N} ratio shown in Figure \ref{fig:HC7N_ratio} eliminates the uncertainty in \ce{H2} by only considering carbon chain emission over the same spatial area. This result also shows that the clumps are likely less than 1 Myr. 

Another factor that could influence our abundance calculations of the carbon chain molecules is the largest angular scale of the VLA.  The largest recoverable angular scale for the VLA in D configuration at K-band is 66 arcseconds \footnote{\url{https://science.nrao.edu/facilities/vla/docs/manuals/oss/performance/resolution}}, which is smaller than some of our sources and could potentially lead to an underestimation of the total carbon chain flux and thus the abundance. An underestimation of the carbon chain flux, however, would lead to a lower abundance of carbon chains relative to \ce{H2} and thus an older age estimate, since the carbon chain abundance decreases to a steady state after $\sim1$ Myr (see Figure \ref{fig:h}). Therefore, any missing flux due to the largest recoverable angular scale does not affect our overall conclusion that these 11 clumps are less than 1 Myr old.

Out of all of the uncertainties considered, the uncertainties in the chemical network appears to have the largest affect on our age determination, but still don't appear to change our overall results. The final conclusion that the detection and abundances of the carbon chain molecules indicate that these 11 sources are less than 1 Myr old appears to be robust against the various sources of uncertainty considered here. 

\subsection{Warm Carbon Chain Chemistry}
Carbon chain molecules can be also produced from the process of WCCC around protostars. WCCC is initiated by \ce{CH4} sublimating from dust grains when temperatures get above $\sim$ 25 K around protostars \citep{2013ChRv..113.8981S}. WCCC, however, tends to be localized ($\sim$ 5000 au) around protostars in star forming regions (e.g. \citealt{Sakai08,2013ChRv..113.8981S}), while our carbon chain observations probe larger scales of dark clumps (on the order of a parsec) and our carbon chain spectra are spatially averaged over the extent of the clumps. \cite{Sakai08} find that the carbon chain molecule CCH, created via WCCC in the core L1527, has an excitation temperature of 12.3 K while the average excitation temperature of CCH in 16 starless cores is 4.9 K. Furthermore, the kinetic temperature of these clumps, measured from \ce{NH3} observations, show that all clumps have a temperature less than 25 K (the temperature required for WCCC). The WCCC emission from L1527 likely comes from a region with a density of $7 \times 10^5 $ cm$^{-3}$ \citep{2013ChRv..113.8981S}, which is an order of magnitude more dense than our dark cloud sources. We conclude that it is unlikely that the carbon chain molecules we detect in these sources are created via WCCC.

It is also possible that carbon chain molecules are created in the outflow cavities of low-mass protostars by photodissociation or shock chemistry \citep{Mendoza18,Zhang18}. \ce{HC5N} was detected in the shock region L1157-B1 with an abundance relative to \ce{H2} of 1.2$\times10^{-9}$ \citep{Mendoza18}. The abundance of \ce{HC5N} in L1157-B1 source suggests that the molecules are formed by chemistry within the shock. It is possible that there are low-mass protostars in our sources with outflow regions that are creating carbon chain molecules. The carbon chain lines in our 11 sources, however, have an LSR velocity that is within 0.5 km/s of that of the parent cloud (see Figure \ref{fig:d}), whereas the \ce{HC5N} in the shock region L1157-B1 is blue-shifted by 3.5 km/s \citep{Mendoza18}. The line widths of our carbon chain lines are also relatively narrow ($\sim$1-2 km/s), which is not what is expected for a shock/outflow region. The line widths of the carbon chain lines from \cite{Mendoza18} are $\sim5$ km/s. With our current data, we ultimately cannot rule out the possibility of carbon chain formation in the formation of outflow regions of low-mass protostars, however, it seems unlikely given the velocities and line widths of the carbon chain lines.

Carbon chain molecules could also be created in outflow cavity walls through other mechanisms separate from shock waves. The cavity wall, created by an outflow from a low-mass protostar, could be illuminated by photons from the protostar, potentially leading to the creation of carbon chain molecules through photochemistry. The current VLA data cannot rule out this possibility given the scale ($\sim$parsecs) and distance of these clumps with the angular resolution of the VLA data. We do, however, find that the exposure to UV radiation does not have a large effect on our age estimates for these sources (see Section 5.2.2).

\subsection{Star Forming Potential of these SMDCs}
All of these clumps have sufficient mass and densities ($n$(H$_2$)=2$\times$10$^4$ cm$^{-3}$) to form high-mass stars \citep{mckee_ostricker2007}. \cite{svoboda19} find that they also meet the mass-radius criteria for HMSF proposed by \cite{Kauffman2010}. Using line width measurements of \ce{NH3}, \cite{svoboda16} calculated the virial parameter for 9 of the clumps in our sample. For clumps with negligible magnetic fields, a virial parameter of $\alpha=1$ represent gravitational virial equilibrium, while a virial parameter of $\alpha \sim 2$ is marginally gravitationally bound \citep{kauffmann13}. The median virial parameter of these source was 0.7, indicating that they are gravitationally bound, neglecting magnetic fields.

These sources lack indicators of HMSF such as \si{70}{\micron} compact regions, \ce{H2O} and \ce{CH3OH} masers, and ultra-compact HII regions. The lack of high-mass stars in these sources could ultimately be for two reasons. The first is that the sources are too young to have formed high-mass stars and the second is that these clumps are inefficient at forming high-mass stars. To answer the question of whether or not these clumps are inefficient at forming high-mass stars, we use a statistical reasoning that it is unlikely that all of the clumps would be inefficient at HMSF given that we measured them all to be young ($\lesssim$ 1.0 Myr old at densities of $n$(\ce{H2})=\num{2e4} cm$^{-3}$). Since, if all of these clumps were inefficient at forming high-mass stars, they could have been any age and we would have expected their ages to be equally distributed over a wide age range. Instead, we find that all of these clumps have similar youthful chemical ages of less than 1 Myr indicated by the presence of carbon chain molecules.

One estimate for the timescale of dissipation of giant molecular clouds by galactic shear in the Milky Way is between 21-60 Myr \citep{Jeffreson18}. We can use this timescale to estimate the probability that all of our clumps are observed to be less than 1 Myr old if they were inefficient at forming high-mass stars. We use the minimum lifetime of a molecular cloud that doesn't form stars and dissipates from galactic shear of 21 Myr, and then assume that if our sources were inefficient at forming high-mass stars, they would be equally distributed in age bins of 1 Myr across the range 0-21 Myr. The estimated probability that the 11 sources are less than 1 Myr old if they are inefficient at forming high-mass stars is $\sim 3\times10^{-15}$ (this corresponds to Gaussian 8-$\sigma$). For this calculation, we do not include the source with the non-detection of carbon chains, however, we note that the non-detection of carbon chains in G23605 could indicate a later evolutionary stage for that source. It is possible that this source will not go on to form high-mass stars and is inefficient at HMSF. The lack of carbon chains seen in this source could indicate that is is at an age greater than 1 Myr.

There is no reason why all the clumps would be young if they are inefficient at forming high-mass stars, since if they were inefficient at HMSF, we would expect them to be evenly distributed across a wide age range up to 21-60 Myr. Thus, we conclude that it is unlikely that all these clumps would be inefficient at high-mass star formation given that all 11 of them with carbon chain detections are less than approximately 1 Myr old at densities of $n$(\ce{H2})=\num{2e4} cm$^{-3}$. This result combined with the physical conditions of these clumps is consistent with these sources not having enough time to form high-mass stars, but still being capable of HMSF at later times. We cannot be certain that all these clumps will all go on to form at least one high-mass star, however, the detection of carbon chain molecules in 11 clumps and corresponding age estimation suggests that it is unlikely that all the clumps are inefficient at forming high-mass stars. This implies that these SMDCs could be representative of the initial conditions of HMSF as these clumps are still at early stages in their evolution. 

The absolute timescale for HMSF is hard to constrain observationally, but current estimates range between $\sim 0.1-1.7$ Myr. A study of 111 massive clumps at different evolutionary stages found a timescale of HMSF (the time required to form a high-mass ($\geq8M_{\odot}$) young stellar object that is bright at mid-IR wavelengths) of 0.41 Myr by modeling the physical properties (mass, luminosity, molecular column density) of the clumps throughout their evolution \citep{sabatini21}. Using \textit{Spitzer} and \textit{Herschel} data of a sample of dense molecular clouds, including clouds in both the starless and star forming phase, \cite{battersby17} estimate the lifetime of the starless phase of molecular clouds to be 0.2-1.7 Myr. This estimated timescale, however, relies on multiple assumptions about the physical properties of the clouds and is thus very uncertain \citep{battersby17}. These estimates of the timescale of HMSF are generally consistent with our conclusion that our 11 clumps are too young to have formed high mass stars, as our estimated maximum chemical ages of $\lesssim$1 Myr is comparable to these estimated timescales of HMSF.

\subsection{CCS as an Early Time Indicator}

While we rely on our measured HC$_5$N and \ce{HC7N} measurements to determine ages for the clumps, the detection of CCS may also support our conclusions that these clumps are young, even though the UMIST models under-predicted the CCS abundance, because CCS has previously been observed in starless cores and is thought to be an early time species that becomes depleted at later evolutionary stages\citep{2013ChRv..113.8981S,Tatematsu2017,Seo19}.
It is thought that molecular clouds take $\sim$1 Myr to reach chemical equilibrium \citep{Yamamoto17}. The high-abundance of CCS in our sources suggest that these sources have not yet reached the evolutionary stage of chemical equilibrium \citep{Yamamoto17}.
To compare the 1 Myr timescale of chemical equilibrium with the dynamical timescale, we estimate the free-fall timescale of these sources. In an ideal super-critical collapse, the free-fall timescale is given as 
\begin{equation}
    t_\mathrm{ff}=\sqrt{\frac{3\pi}{32G\rho_\mathrm{gas}}}\simeq 1.38\times10^6(\frac{n_{\ce{H2}}}{\SI{e3}{cm^{-3}}})^{-1/2} \ \left[\si{yr}\right]
\end{equation}
where $n_{H_2}$ is the number density of \ce{H2}. Using equation 10 and the mean density we calculate for our SMDCs, $n(\ce{H2})=\SI{2e4}{cm^{-3}}$, we estimate the free fall timescale of these sources to be $\approx\!\num{4.4e5}$ years. The speed of collapse of molecular clouds is generally observed to be $20-50\%$ less than the predicted free-fall speed because of magnetic field and turbulence support against gravitational collapse \citep{wyrowski2012,Infalls,Feng}.
This correction gives a contraction timescale of  $t_\mathrm{cont}\approx10^6$ years. 

This free-fall timescale is comparable to the timescale of chemical equilibrium, suggesting that these sources have either not yet reached their free-fall timescale or are just reaching this timescale of $\num{4.4e5}$ to $\num{1e6}$ years. Given that 10 out of 11 of the sources with carbon chains have previous detections of low-mass protostars \citep{svoboda19}, it is possible that these 10 clumps are closer to the 0.1-1 Myr end of the estimate age distribution (shown in Figure \ref{fig:h} and Figure \ref{fig:HC7N_ratio}) and that some of the denser regions in the clump have reached their free-fall timescale. The calculated free-fall timescale is averaged over the entire clump from the estimated the \ce{H2 column density}, so there may be denser regions of the clump that have collapsed to form the first set of protostars in these sources. 

The observed protostars from \cite{svoboda19} are likely Class 0 given how embedded they are in their clumps and the sizes of their outflows and they may represent the first protostars to form in these clumps, and not the entire population of protostars that could form from these clumps. Based on the size of the outflows and their velocities, \cite{svoboda19} estimated an age of these protostars of $\sim$50 kyr. The ``half-life" of Class 0 protostars (the timescale over which half the protostars in a class have a shorter or longer lifetime ) is $\sim$50 kyr and that of the Class II phase is 2 Myr \citep{Evans03,Kristentsen18}. We therefore may not expect the entire possible population of stars to have formed and evolve on the same timescale from these clumps. It is then also unlikely that many of these protostars have evolved past the Class 0 protostar phase at the age estimates we obtain from the carbon chain chemistry and based on the age estimates from \citep{svoboda19}. All of this could then explain why some protostars have formed in these clumps, even though the carbon chain chemistry analysis reveals that the age of the clumps is likely comparable to their clump averaged free-fall timescales.

\subsection{Comparisons with Previous Work}
The age estimations we measure for our SMDC sources are generally consistent with ages of other SMDC measured by \cite{Feng}. \cite{Feng} conducted a line imaging survey of infrared dark clouds less than \SI{5}{kpc} away that have similar physical properties to our sources. \cite{Feng} determines the ages of their SMDC sources by measuring CO depletion and \ce{HCO+} deuterium fractionation and comparing to a dark cloud chemistry model. The ages determined by \cite{Feng} are \SI{8e4}{yr}, which is less than the maximum chemical age we find of 1 Myr for our sources. We do not have the same precision with the chemical models as \cite{Feng} given our age determination only comes from carbon chain abundance measurements, so we cannot be more precise than saying these clumps are likely less than 1 Myr old at densities of $2\times10^{4}$ cm$^{-3}$. \cite{Feng} finds that the chemical ages of their sources are comparable to their free-fall timescale and smaller than their contraction timescale, which indicates that their sources are dynamically young as well. 
 
Our measured ages are also generally consistent with the chemical timescales presented in \cite{Gerner}. \cite{Gerner} conducts a survey of 59 different high-mass star forming regions at different evolutionary stages and suggests that the timescale for the evolutionary sequence of high-mass star formation (going from a starless infrared dark cloud to an ultra compact HII region) is on the order of 10$^5$ years, which is the same order of magnitude as our maximum chemical ages of these 11 sources. The consistency of our age estimations with \cite{Feng} and \cite{Gerner} is further support that our 11 SMDC sources are in the evolutionary stage preceding the formation of high-mass stars.

\section{Conclusion}\label{sec:Conclusion}
We present VLA molecular spectral line observations of carbon chain molecules towards 12 SMDCs that were previously identified with the Bolocam Galactic Plane Survey. These sources have a median mass of 790 $M_{\odot}$, greater than the estimated mass required to form a single high-mass ($\geq 8 M_{\odot}$) star. We look for HC$_5$N, CCS and HC$_7$N in each of the 12 sources. The observations of carbon chains in these molecular cloud clumps probe the evolutionary stage of these clumps, as the carbon chains are expected to be an early-time species. Our main findings from these observations are:

\begin{itemize}
\item We detect the carbon chain molecules HC$_5$N and CCS towards 11 out of 12 of \SI{70}{\um}-dark clumps in the sample.  
\item We do not have any clear 3$\sigma$ HC$_7$N detections, but averaging all the spectra from both transitions from all 11 sources with carbon chain detections shows that there is a HC$_7$N detection on average in these sources.

\item From comparing our measured abundances of HC$_5$N to dark cloud chemistry models, we find that these 11 sources are most likely less than 1 Myr old.
\item In comparing our measured \ce{HC5N} to \ce{HC7N} ratio to the ratio predicted by the UMIST 13 dark cloud chemistry models, we find that our sources are less than $\sim $1 Myr old, which is consistent with our age estimation from the \ce{HC5N} abundances.
\item Since these 11 clumps were found to be young, we conclude that it is unlikely that they all would be inefficient at forming high-mass stars. If these sources were inefficient at forming high-mass stars, we would expect to see them evenly distributed across a wide age range (up to $\sim$20-60 Myr). We, however, observe them to all be less than $\sim$1 Myr old at densities of $n(\ce{H2})\approx\SI{2e4}{cm^{-3}}$.  Thus, given that these sources meet the criteria for HMSF, they could go on to form high-mass stars at later times.

\end{itemize}

\section{Acknowledgments}
 KW and BES acknowledge support by the National Science Foundation Research Experiences for Undergraduates (REU) program for supporting this work. KW and BES thank the referees for their comments which helped improve this manuscript. Kadin Worthen was a
 summer student at the National Radio Astronomy Observatory. KW acknowledges support from the NASA FINESST program. This work is supported by the National Aeronautics and Space Administration under Grant No.
80NSSC22K1752 issued through the Mission Directorate. The National Radio Astronomy Observatory and Green Bank Observatory are facilities of the U.S.
National Science Foundation operated under cooperative agreement by Associated Universities, Inc.

\facilities{
    VLA
}

\software{
This research has made use of the following software projects:
    \href{https://astropy.org/}{Astropy} \citep{astropy18},
    \href{https://matplotlib.org/}{Matplotlib} \citep{matplotlib07},
    \href{http://www.numpy.org/}{NumPy} and \href{https://scipy.org/}{SciPy} \citep{numpy07},
    \href{https://pandas.pydata.org/}{Pandas} \citep{pandas10},
    \href{https://ipython.org/}{IPython} \citep{ipython07},
    \href{https://casa.nrao.edu/}{CASA} \citep{casa07},
    and
    the NASA's Astrophysics Data System.
}
\appendix
\section{\ce{HC5N} moment 0 maps}
Here we show the moment 0 maps of \ce{HC5N} for the remaining sources with carbon chain detections not shown in main the body of the manuscript (see Figure \ref{fig:hc5n_mom_maps}). These moment 0 maps are computed before the primary beam correction to reduce noise on the edges of the images. The moment 0 maps are computed only over the velocities where the line is present as shown in Figure \ref{fig:d}. The \ce{HC5N} emission is detected clearly in the moment 0 maps of G23297, G28565, G9558, and G30912. The \ce{NH3} contour used for the carbon chain extraction is  overplotted as red lines on the \ce{HC5N} moment 0 maps in Figure \ref{fig:hc5n_mom_maps} to check that the \ce{HC5N} and \ce{NH3} are cospatial. 
\begin{figure*}[!htbp]
    \centering
    \includegraphics[scale=0.55]{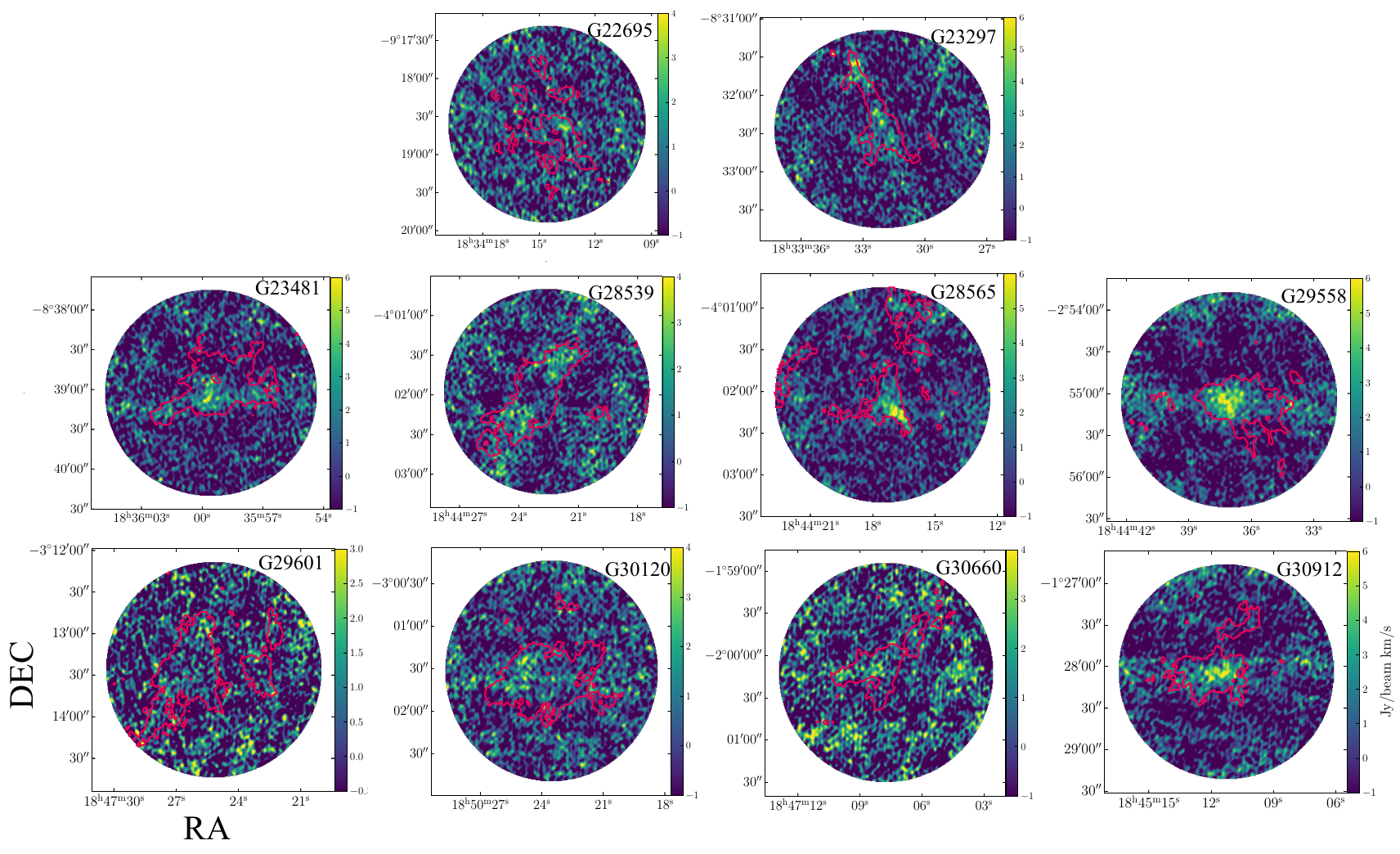}
    \caption{ \ce{HC5N} moment 0 maps of the remaining 10 sources with carbon chain detections (excluding G24051, which was shown in the main text) with \ce{NH3} contour overlaid. This is the same \ce{NH3} mask that was used for the spectral extraction. }
    \label{fig:hc5n_mom_maps}
\end{figure*}


\bibliographystyle{aasjournal}
\bibliography{mybib}

\newpage






\end{document}

%% file: commands.tex
\DeclareSIUnit{\jy}{Jy}
\DeclareSIUnit{\beam}{beam}
\DeclareSIUnit{\kms}{\kilo\meter\per\second}